\documentclass[a4paper,fleqn,usenatbib]{mnras}
\usepackage{newtxtext,newtxmath}
\usepackage[T1]{fontenc}
\usepackage{ae,aecompl}
\usepackage{graphicx, subfloat}	
\usepackage[graphicx]{realboxes}% Including figure files
\usepackage{amsmath}	% Advanced maths commands
\usepackage{float}
\usepackage{subfloat}
\usepackage{caption}
\usepackage{adjustbox}
\usepackage[dvipsnames]{xcolor}
\usepackage{mathtools}
\usepackage{rotating}
\usepackage{pdflscape}
\usepackage{graphicx}
\usepackage{pdflscape}
\usepackage{multicol}
\usepackage[T1]{fontenc}
\usepackage{rotating}
\usepackage{subcaption}
\usepackage[graphicx]{realboxes}
\usepackage{wrapfig}
\usepackage[flushleft]{threeparttable} % http://ctan.org/pkg/threeparttable
\usepackage{booktabs,caption}
\hypersetup{draft}

\title[Outflows in the Cygnus X Region]{Characterizing Outflows in the Cygnus X Region}

\author[Deb, Kothes, \& Rosolowsky]{
Soumen Deb,$^{1}$\thanks{E-mail: sdeb@ualberta.ca}
Roland Kothes,$^{2,1}$
Erik Rosolowsky,$^{1}$
\\
$^{1}$Department of Physics, University of Alberta, Edmonton, Alberta, Canada T6G 2E1\\
$^{2}$Dominion Radio Astrophysical Observatory, Herzberg Programs in Astronomy \& Astrophysics, National Research Council Canada, \\~~~P.O. Box 248, Penticton, BC V2A 6J9, Canada \\
}

\date{Accepted XXX. Received YYY; in original form ZZZ}

\pubyear{2018}
\defcitealias{gott12}{G12}
\defcitealias{deb18}{D18}
\begin{document}
\label{firstpage}
\pagerange{\pageref{firstpage}--\pageref{lastpage}}
\maketitle

\begin{abstract}
% Cygnus X is an active star-forming region in the local spiral arm and a massive reserve of molecular gas. The presence of low-velocity,  high-momentum molecular outflows is an inherent characteristic of the embedded prenatal phase of stars. 
In this paper, we perform an analysis of 13 outflows in the Cygnus X star-forming region. We use the James Clerk Maxwell Telescope observations of $^{13}$CO(3-2) and C$^{18}$O(3-2) molecular emission lines combined with archival $^{12}$CO(3-2) data.  Using these new observations, we measure the mechanical properties of the outflows,    and identify the associated protostars, finding their properties consistent with previous surveys of outflows throughout the Milky Way. Finally, we develop and test a method to measure the same properties using the existing $^{12}$CO(3-2) line data alone, finding the properties agree to within a factor of 2.
% A comparison of values reveals our model based on only the $^{12}$CO(3-2) line is reliable for estimating outflow properties without multi-tracer data.
\end{abstract}


%%%%%%%%%%%%%%%%%%%%%%%%%%%%%%%%%%%%%%%%%%%%%%%%%%

%%%%%%%%%%%%%%%%% BODY OF PAPER %%%%%%%%%%%%%%%%%

\section{Introduction}
Star formation shapes structure and evolution of a galaxy by consuming gas and injecting feedback into the interstellar medium (ISM). A significant amount of feedback comes from the momentum and energy from the winds of massive stars but also some fraction of feedback comes from protostellar outflows.

All accreting astronomical objects tend to have bipolar outflows or collimated jets, resulting from the interaction between the gravitational potential of the central rotating object and the magneto-centrifugal potential arising from the accretion disk \citep{krum, bally16}. Accreting neutron stars, quasars, active galactic nuclei, and young stellar objects (YSO) all show bipolar outflows at some point in their lives. While the bipolarity, degree of collimation, and morphology of these outflows are similar regardless of their origin, some outflow properties depend on the central object. For example, for outflows generated by protostars, the ejecta velocities can vary from 1 to 100 km~s$^{-1}$, whereas neutron stars can produce outflow velocities at a significant fraction of the speed of light. 

Outflows set in as soon as the accretion disks around collapsing protostars are formed. The outflows associated with young stellar objects (protostars) provide useful information about the evolutionary stages of forming stars as well as the condition of the parent clouds, since the size, velocity, mass and momentum of the ejecta depend on the generating YSO (protostar) and the cloud environment \citep{bally16}. 

Most YSOs show two components: a high-speed, relatively low-mass collimated jet of atomic or ionized matter, and a wide angled, slow-moving, massive molecular component. The bipolar atomic/ionized jet is emitted orthogonal to the plane of the accretion disk reaching large distances. The molecular component appears more closely connected to the rotating core. Outflows inject mass and momentum into the protostellar environment in opposite directions perpendicular to the plane of accretion, and the mass injection rate increases with the accretion rate \citep{eller13}. During the early stages of protostellar evolution, molecular outflows are the dominant sources of momentum and energy injection to the natal cloud. Additionally, the physical characteristics continuously evolve with the YSOs. In the early stages of class 0 YSOs, outflows are predominantly molecular and become progressively more atomic and ionized with increasing velocities as the YSOs evolve into class I. 
% Redundant with below
% In molecular clouds hosting low-mass star formation, the momentum and energy from protostellar outflows make significant contributions of feedback.  This feedback has been proposed as a mechanism to shape evolution of the molecular clouds and the resulting stellar initial mass function by providing turbulence and self-regulating star formation \citep{elm04,bally16}. 
% For these reasons, the study of outflows and their feedback is particularly important in the context of star formation.

Because of the multiple ionization states for outflowing material, several tracers are required for revealing all the different features of outflows. 
%The observed morphology and physical properties of proto-stellar outflows depend on the tracer used. 
The atomic and ionized components of the jet are observed with radio and x-ray continuum emission and the (semi-) forbidden line transitions of atomic species in the optical and UV.  While the molecular component can be traced through the infrared lines of H$_2$, the low-$J$ rotational transition lines of CO molecules are the most commonly used tracers because of their brightness and their observability. The lines are bright because of the relatively high fractional abundance of CO and the high likelihood of collisions with H$_2$ and He that populate the low-$J$ states. The low $J$ transitions can be observed in the millimetre/submillimetre regimes with ground-based facilities.
% emission for the jets, near infrared lines of hydrogen, optical forbidden transition lines of sulfur, nitrogen, oxygen and iron for the ionized and atomic matter, even UV lines. For the past few decades, the low-$J$ rotational transition lines of CO molecules are the most commonly used tracer for observing the molecular component of the outflows, which is the dominant source of momentum and energy injection during their early evolutionary stages into the low-mass star-forming regions of the ISM. 
% This is primarily because of the abundance of CO lines via collisional interaction with hydrogen and helium in typical star-forming molecular clouds and their observational ease with ground based telescopes. 
In addition, high spectral resolution observations can measure the Doppler broadening of the spectral line profile, which reveals the characteristic line wing features in outflows. These line wings extend $10$ to $100$ km~s$^{-1}$ from the line centre. 
% \citet{wil70} were among the first authors revealing such line broadening using $^{12}$CO (1-0) line emission from molecular clouds in the Orion nebula.  

With molecular spectroscopy we can measure several properties of the outflows using bipolar wings.  The standard properties inferred are size, morphology, mass, momentum, energy, and mechanical luminosity. By comparing these quantities with the proto-stellar luminosity, we constrain the accretion time, the efficiency of outflow launching, and the momentum and energy injection rates into the ISM. For low-mass cores this feedback is conjectured to play a significant role in providing turbulence and maintaining virial balance against the core gravitational energy in collapsing clouds. For massive cores, outflow feedback can potentially disrupt and shred the cloud \citep{bally16}. However, the impact of outflow feedback and its coupling to the parent clouds remains uncertain. Some studies have argued that outflows provide a minimal contribution of feedback \citep[e.g.,][]{hansen12} and may not be effective at driving local cloud turbulence \citep{swift08,duartecabral12,drabek16}.  However, these estimates rely on careful characterization of outflow properties over a large region within the host molecular cloud. The impact of outflows could be larger than previously considered. \citet{Dun} argued that outflow mass, momentum estimated from low-$J$ CO lines only provide lower limits on those quantities given the standard assumptions about opacity and line excitation.

Most outflow studies have focused on the nearest star-forming regions, which are mostly relatively quiescent.  The Orion molecular cloud is the nearest site of high-mass star formation and remains the case study for outflows and feedback from newly forming O and B stars \citep{bally16}. However, on Galactic scales, Orion is a relatively small molecular cloud and more distant regions contain larger molecular clouds and a wealth of outflow activity. In this work, we study the Cygnus~X region, a massive molecular cloud complex associated with the spiral arm.  Cygnus~X is the most active star-forming region within 2~kpc and shows a range of outflow behaviour across the region.

% In the Milky Way galaxy there are a number of gas-rich complexes that are sites for large clusters of young massive stars. There are hundreds of protostellar outflows in the solar neighborhood. Most of the outflow studies have focused on nearest clouds and particularly the Orion Nebula. However, Cygnus X is a nearby giant molecular cloud (GMC) complex showing vigorous high mass star formation. Cygnus X is located in the local spiral arm or the Orion Spur, and the fact that it is the most active star-forming region within 2 kpc radius from the Sun makes it an interesting target for a study of galactic star formation process. 

This work is a follow-up of a survey of the Cygnus~X region in $^{12}\mathrm{CO}(3-2)$ emission made with the James Clerk Maxwell telescope (JCMT) by \citet[][hereafter \citetalias{gott12}]{gott12}. The \citetalias{gott12} work identified 47 molecular outflows in the $^{12}\mathrm{CO}$ emission.  In this work, we present $^{13}\mathrm{CO}(3-2)$ and $\mathrm{C^{18}O}(3-2)$ observations of 13 of these outflows to measure their properties. This work extends the analysis presented in \citet[][hereafter \citetalias{deb18}]{deb18}, which studied one object in detail in the context of triggered star formation. Here, we use standard approaches to measure the outflow properties for the combined sample of 13 outflows.  

In addition, we also develop a procedure to measure the properties of outflows using only the $^{12}\mathrm{CO}(3-2)$ emission line.  This method is motivated because we have carried out a wide area survey of the Cygnus X region in $^{12}\mathrm{CO}(3-2)$ emission using the JCMT that will be presented in forthcoming work (Deb et al. in preparation).  As part of that survey, we have identified hundreds of protostellar outflows.  Measuring the properties using multiple CO tracers of those outflows would require a heavy investment of telescope time.  Hence, validating the methods for a single-tracer measurement of outflow properties is important for studying outflows in the context of feedback.

% We have an existing $^{12}$CO(3-2) large survey data in which we have visibly estimated over a hundred outflows that we intend to study in a follow-up paper. However, in order to estimate outflow properties as done in \citet{deb18} we would require $^{13}$CO(3-2) as well as C$^{18}$O(3-2) rotational transition line data. This in turn implies the requirement of a significant amount of dedicated telescope time with the JCMT which may not be feasible. 
% Thus, we propose a prescription for extrpolating the same dynamical properties of outflows only using the existing $^{12}$CO(3-2) line data. Since we have all three lines available for these 13 outflows we compare out estimation based on them against that based on the extrapolation model for its validity for use in the follow-up work mentioned above.

Specifically, we detail our observational techniques and data extraction in Section \ref{results}. In Section \ref{allco}, we discuss the properties of all three CO rotational lines and assuming a constant excitation temperature among all species we determine the optical depths and column densities of the optically thin lines as functions of position and velocity offsets from the line centre. Section \ref{wings} shows how we calculate mass, momentum, and energy of the molecular outflows using all three tracers following a similar approach as \citetalias{deb18}. Finally, in section \ref{12to13} we present a model for extracting outflow properties from $^{12}$CO(3-2) line alone and compare the results to the three-line estimates.

\section{Observations} \label{obs}
Here, we have used rotational transition lines $^{13}$CO(3-2) and C$^{18}$O(3-2) observed in the bands centered at 330.58 and 329.33 GHz, respectively, using the JCMT at the summit of Mauna Kea in Hawai'i, using the Heterodyne Array Receiver Program (HARP) instrument and the Auto Correlation Spectral Imaging System (ACSIS) spectrometer (see also \citetalias{deb18}). In Table \ref{obs_compare}, we summarize some of the observational details of the 13 outflow sources, including project codes, weather bands and mean atmospheric opacity values at 225 GHz during the observational runs (March 2010, and July 2011) at the JCMT. Most sources were observed using ``jiggle'' mapping but the largest source was observed using a raster map.  We configured the receivers and ACSIS correlator to provide 61 kHz spectral resolution in simultaneous observations across the two spectral lines.  

The 13 outflows presented here were part of the larger sample of outflows identified in \citetalias{gott12}. For this project the brightest outflows from G12 were selected. While we planned to observe more outflows, we only obtained data on these 13 targets based on the constraints set of observational feasibility (telescope scheduling and weather).  Thus, our actual sample is not designed to statistically represent the parent outflow population.

For data reduction, we used the observatory-maintained {\sc starlink} software package \citep{STARL} and the standard reduction and calibration recipes developed for the JCMT. The observatory provides calibrated data on the $T^{*}_\mathrm{A}$ scale (antenna temperature corrected for atmospheric opacity, but not for source-beam coupling).  We convert the data to the main beam temperature scale by assuming a beam efficiency based on observatory recommendations of $\eta_\mathrm{MB}=0.64$\footnote{\url{https://www.eaobservatory.org/jcmt/instrumentation/heterodyne/harp/}} and setting $T_\mathrm{MB} = T_A^*/\eta_\mathrm{MB}$. We grid these data into position-position-velocity spectral line data cube with a beam size of $14.6''$ (pixel size of $7.3''$) and velocity resolution of 0.055 km~s$^{-1}$. The central spatial coordinates of each cube are shown in Table \ref{obs_compare} (refer to \citetalias{deb18} for more details). For each position, we defined an emission-free region of the baseline by-eye and then subtracted a linear baseline.  Additionally, we have archival $^{12}$CO(3-2) line data from \citetalias{gott12}, which we re-sampled and aligned to match the same coordinate grid as the $^{13}$CO(3-2) and C$^{18}$O~(3-2) data. The median values of RMS noise in the final  $^{13}$CO(3-2) and C$^{18}$O(3-2) data cubes are 0.31 K and 0.38 K in 0.055 km s$^{-1}$ channels, respectively.  The noise values for the archival $^{12}$CO(3-2) data at the same velocity resolution are larger (typically 0.4 to 0.8 K) but this line is always strongly detected. All three lines are detected at $>5\sigma$ at some position in each of the targets.
 
\begin{center}
    \begin{figure*}
               \includegraphics[width=\textwidth]{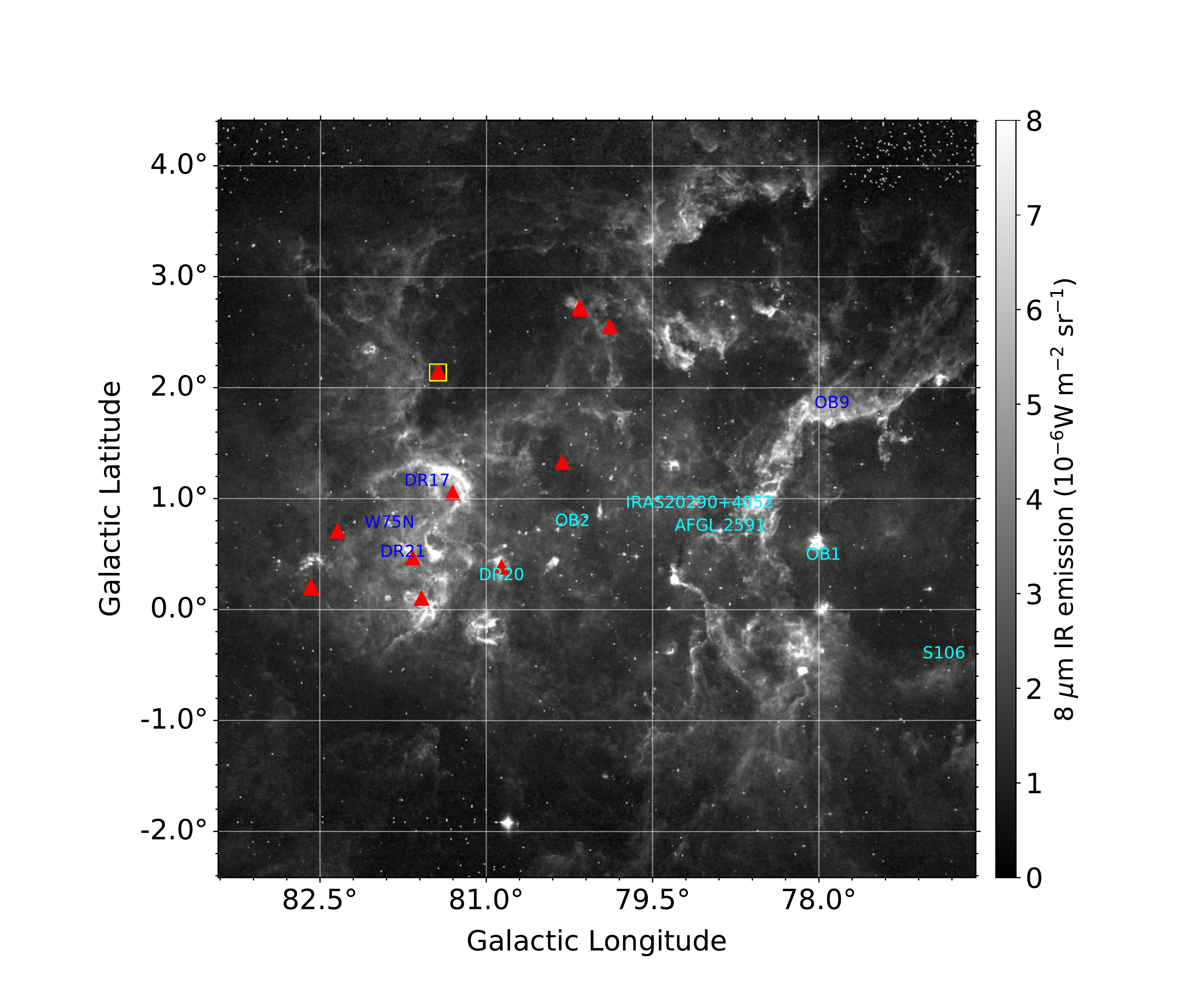}
               \caption{\label{cygx_8mu} Dust thermal emission at 8 $\mu$m reveals molecular clouds in Cygnus X, with the major star-forming regions labeled (blue and cyan). Locations of the outflows discussed in this work are marked with red triangles. The yellow square denotes the location of the cometary feature discussed in \citet{deb18}.}
    \end{figure*} 
 \end{center}

The locations of the 13 observed outflows are shown in Figure \ref{cygx_8mu} with a background of 8 $\mu$m PAH emission, which highlights the regions of star formation in Cygnus X \citep{croc11,peet04}. The large cavities in the 8~$\mu$m emission surround regions where molecular gas was destroyed by newly formed stars giving the popular ``Swiss cheese'' appearance \citep{Bal99}.  The observed outflows are all located near the DR21 region, with eight in the active star-forming region and five outflows in satellite clouds, including the region studied in \citetalias{deb18}.

\begin{figure}
    %\centering
    % \begin{subfigure}[b]{0.47\textwidth}
    %       \centering
\includegraphics[width=0.49\textwidth]{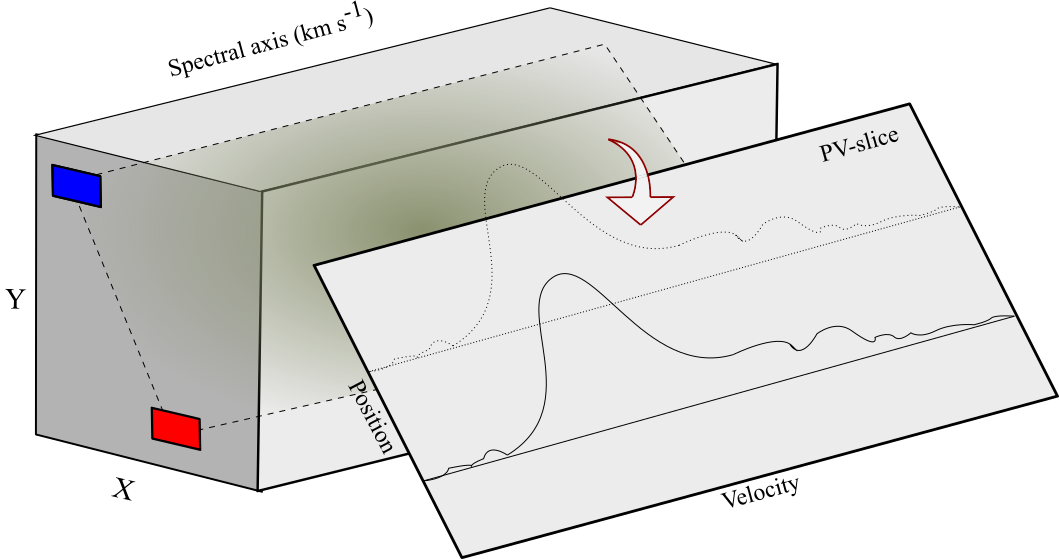}
    \caption{A position-velocity (PV) slice out of a data cube. The  $x{-}y$ plane defines the plane of sight. The third axis is for frequency or equivalently velocity, along which spectral line profiles at each spatial pixel along the $x{-}y$ line as shown in the PV-slice. \label{pv}}%The plot highlights the presence of two outflows highlighted in yellow.
\end{figure}

\section{Results} \label{results}

Here, we present the observations of the individual outflows and describe how we measure their physical properties.

\subsection{Atlas of Outflows}

\begin{figure*}
    \centering
    \includegraphics[width=\textwidth]{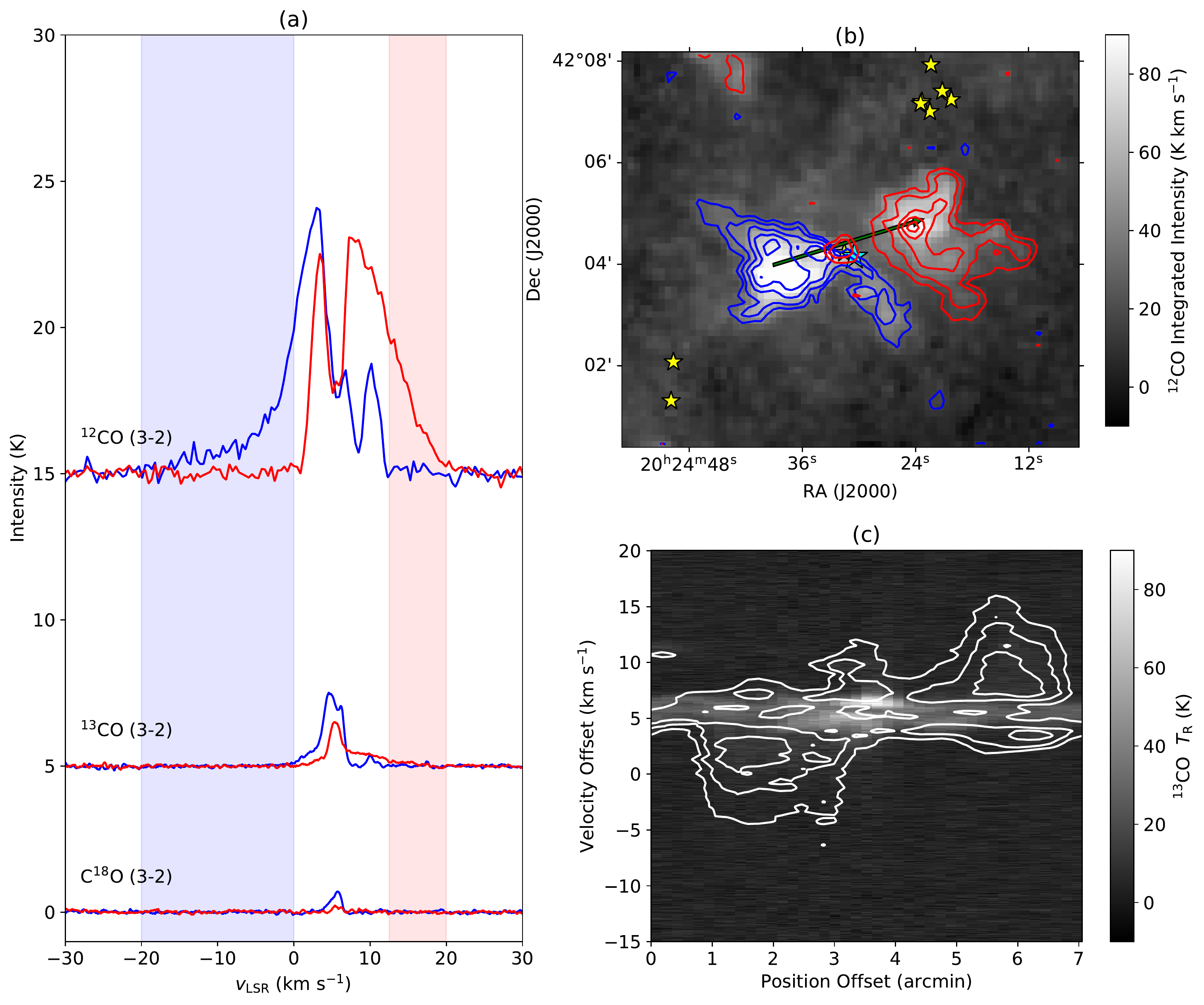}
    \caption{Outflow G79.886+2.552 : (a)  Average spectral intensity in blue- and redshifted outflow regions are shown in $^{12}$CO(3-2) (offset $+15$~K), $^{13}$CO(3-2) (offset $+5$~K), and C$^{18}$O(3-2) lines. The wing feature is present in $^{12}$CO(3-2) line, which is self-absorbed in the line centre caused by the foreground Cygnus Rift. (b) Integrated intensity of $^{12}$CO(3-2) line emission highlights the spatial distribution of molecular gas. Red and blue contours represent the red- and blueshifted wings, plotted over the background of total emission (gray-scale). Blue and red contours are obtained by integrating over velocity ranges of $v=-20$ to $0\mathrm{~km~s}^{-1}$ and $v=12.5$ to $20\mathrm{~km~s}^{-1}$ respectively. Contour lines are drawn at levels (5, 10, 20, 30, 40) K~km~s$^{-1}$ and (4, 8, 15, 20, 25) K~km~s$^{-1}$ respectively. Yellow stars indicate protostars in \citet{kry} catalouge, with the driving IR source marked in cyan. (c) Spatial and spectral distribution of outflowing gas along the PV-slice marked by the green arrow in panel (b). Contours are drawn at levels (3, 5, 7.5, 9.5, 11) K.}
    \label{of1}
\end{figure*}

Our primary data product is a multi-tracer atlas of these 13 outflows. In Figure \ref{of1} we show one of the molecular outflows, G79.886+2.552, from three complementary perspectives. We have included similar three-panel figures for the rest of the 12 sources in a supplemental document that is available online.

By eye, we extract a position-velocity (PV) slice from the data cube (Figure \ref{pv}) that is centred on middle of the outflow and oriented so that the slice goes through the brightest part of the red- and blueshifted outflow lobes.   We extract the property outflows from this PV slice. The PV-slice is one beam ($14.6''$) in width and the emission is spatially averaged perpendicular to the slice direction. We experimented with changing the slice widths but found that the results were most stable for the chosen width, acting as a compromise between including all emission from the outflow and including background emission from the molecular cloud.

The first panel of the atlas (Panel a) displays the spectrum for each of the three CO isotopologues averaged over the red- and blue-shifted sides of the PV slice as the red and blue curves, respectively.  The spectrum shows the contrasts in the different line structure of the three species.  The strong wing features are visible in the high opacity $^{12}$CO(3-2) emission but the optically thin C$^{18}$O(3-2) emission is symmetric and useful for determining the line centre. The shaded bands in blue and red mark the regions we identify, again by eye, as belonging to the blue- and redshifted wings.  We give the values for these boundaries in Table \ref{obs_compare}.

Panel (b) displays the integrated intensity maps of the $^{12}$CO(3-2) emission, which reveal the spatial distribution (size and morphology) of the outflowing molecular gas. Red and blue contours represent the red- and blueshifted wings of the outflow, plotted over the background of total emission (gray-scale). The gray-scale shows the integration over the entire spectral line, but the blue and red contour sets indicate emission over the velocity ranges indicated in panel (a). Yellow stars show the positions of protostellar sources in the region according to the catalogue of \citet{kry}, which was generated from {\it Spitzer}-IRAC survey of the region. We have identified the infrared source that is driving the outflow, marked by a cyan star, by finding the protostar that best matches the position of the centre of the outflow. 

Finally, in panel (c) we display the PV slice for the outflow.  This panel shows the spatially-averaged contour lines of $^{12}\mathrm{CO}(3-2)$ emission along the PV-slice against the background of spatially-averaged $^{13}\mathrm{CO}(3-2)$ emission. The velocity offsets distributed over position offsets indicate the strength of bipolarity in the outflowing gas.

\begin{table*}
\centering
    %\centering
    \captionof{table}{\large{Observational Summary}. The Project ID is the designation from the JCMT. The last five columns give ranges for the blue- and redshifted wings of the outflow and the line centre.} 
    \begin{tabular}{cccccccccc}
     \hline
   Outflow & RA & Dec & Proj. & Atm. & Min.\ blue & Max.\ blue & Line centre & Min.\ red & Max.\ red  \\
        & (J2000) & (J2000) & ID &  Opacity & Vel. & vel. & vel.& vel. & vel.\\
        & & & &  @225GHz & (km s$^{-1}$) & (km s$^{-1}$) & (km s$^{-1}$)&(km s$^{-1}$) &(km s$^{-1}$)\\
         \hline
         \vspace{5mm}
          G79.886+2.552 & $20^{\rm h}24^{\rm m}31.6^{\rm s}$ &$+42^{\circ}04^{'}20.0^{''}$ & M10AC12  & 0.070 &-20 & 0 & 6.3 & 12 &+20\\
          \vspace{5mm}
          G81.435+2.147 &  $20^{\rm h}31^{\rm m}12.5^{\rm s}$  &$+43^{\circ}05^{'}42.0^{''}$ & M10AC12 & 0.069 &-16 & -5 & -2.8 & 12.5 &+15  \\
          \vspace{5mm}
          G81.424+2.140 &  $20^{\rm h}31^{\rm m}12.3^{\rm s}$  &$+43^{\circ}04^{'}53.0^{''}$ & M10AC12 & 0.069 &-14 & -6 & -3.1 & -1.5 & +10 \\
          \vspace{5mm}
          G81.302+1.055 & $20^{\rm h}35^{\rm m}33.5^{\rm s}$  &$+42^{\circ}20^{'}17.0^{''}$ & M10AC12 & 0.083 & +8 & 14.5 &15.4 &17 &+24 \\
          \vspace{5mm}
          G80.314+1.330 &  $20^{\rm h}31^{\rm m}12.3^{\rm s}$  &$+41^{\circ}42^{'}30.0^{''}$ & M11AC10 & 0.062 &-40 & -33.5 &-32.2 &-29.5 &-27 \\
          \vspace{5mm}
          G80.862+0.385 & $20^{\rm h}37^{\rm m}00.6^{\rm s}$  &$+41^{\circ}35^{'}00.0^{''}$ & M10AC12 & 0.060&-15 &-5 & -1.8 & 0 &+8\\
          \vspace{5mm}
          G81.663+0.468 & $20^{\rm h}39^{\rm m}15.9^{\rm s}$  &$+42^{\circ}16^{'}15.0^{''}$ & M10AC12 & 0.075& +10 & 16.5& 19.3 & 23 &+44\\
          \vspace{5mm}
          G81.551+0.098 & $20^{\rm h}40^{\rm m}28.7^{\rm s}$  &$+41^{\circ}57^{'}21.0^{''}$ & M10AC12 & 0.061&-14.5 & -8.5 &-6&-4.5 &+1.8 \\
          \vspace{5mm}
          G81.582+0.104 &$20^{\rm h}40^{\rm m}33.3^{\rm s}$  &$+41^{\circ}59^{'}05.0^{''}$& M10AC12 & 0.082&-15 & -8.5 &-6.2& -4.5 & +2 \\
          \vspace{5mm}
          G82.581+0.203 & $20^{\rm h}43^{\rm m}27.8^{\rm s}$  &$+42^{\circ}49^{'}58.0^{''}$ &  M10AC12 & 0.120 &-10.5 & 6.5 &10.2 & 15.5 &+32 \\
          \vspace{5mm}
          G82.571+0.194 & $20^{\rm h}43^{\rm m}27.9^{\rm s}$  &$+42^{\circ}49^{'}11.0^{''}$ &  M10AC12 & 0.120 &-4 & 7.5 & 11.0 & 13.5 & +23 \\
          \vspace{5mm}
          G80.158+2.727 & $20^{\rm h}24^{\rm m}35.7^{\rm s}$  &$+42^{\circ}23^{'}41.0^{''}$  & M11AC10 & 0.066 & -25  & 0.5 & 4.5 & 10 & +21  \\
          \vspace{5mm}
          G80.149+2.710 & $20^{\rm h}24^{\rm m}38.6^{\rm s}$  &$+42^{\circ}22^{'}42.0^{''}$ &  M11AC10 & 0.066 & -3 & 3 & 5.0 & 6 & +12 \\
         \hline
    \end{tabular}
      %\begin{tablenotes}
  %\item[*] $^{13}$CO(3-2)
  %\end{tablenotes}
\label{obs_compare}
\end{table*}

\subsection{Distances}\label{sec:distances}

We determine the distances to each outflow based on their mean line-of-sight velocity. Referring to \cite{rygl12} and \cite{gott12} we associate the outflows here with four different major star-forming regions in Cygnus X. The range of local standard of rest velocities ($\varv_{\rm LSR}$) of the 13 outflows are included in Table \ref{obs_compare}. Using water masers \cite{rygl12} determined the average proper motion velocities of the two star-forming regions W 75N and DR 21 to be 9 km s$^{-1}$ and -3 km s$^{-1}$, along with their parallax distances. Hence we consider an outflow with a slightly positive velocity ($0 < \varv_{\rm LSR}/(\mathrm{km~s^{-1}})<8$) towards Cygnus X to be at the same distance as Cygnus Rift, which is at a mean distance of 650 pc from the sun \citep{gott12} and one with a low negative velocity ( $-10<\varv_{\rm LSR}/(\mathrm{km~s^{-1}})<0$  to be associated with DR 21, at $1.5$ kpc. We associate outflows with positive LSR velocities ( $\varv_{\rm LSR}/(\mathrm{km~s^{-1}})>8$) with W 75N, at $1.3$ kpc. Outflows G81.435+2.147 and G81.435+2.147 are part of the cometary feature mentioned in \cite{deb18} and are being irradiated by Cygnus OB2 hence we assumed a distance of 1.4 kpc for them.   

\subsection{CO line emission: Column Density}\label{allco}

To measure the physical properties of the outflows, we extend the work of \citetalias{deb18} to determine the outflow column density as a function of velocity from the CO lines.  The column density estimates are controlled by the opacity of the underlying tracer \citep{Oos}, so we measure the opacity of the spectral line as a function of position using the three molecular rotational transitions ($^{12}$CO(3-2), $^{13}$CO(3-2) and C$^{18}$O(3-2)). Using the radiative transfer equation and the emission model from \citet{mangum}, we express the spectral line emission in terms of radiation temperature as a function of optical depth $\tau_\nu$, $T_R=[J_{\nu}(T_{\rm b})-J_{\nu}(T_{\mathrm bg})](1-e^{-\tau_{\nu}})$, where $J_{\nu}(T)=\frac{c^2}{2k\nu^2}B_{\nu}(T)$ and $T_{\mathrm bg}$ is the constant background temperature, taken to be the cosmic microwave background ($T_\mathrm{bg} \approx 2.73$~K).

We assume local thermodynamic equilibrium (LTE) in the molecular gas, and use a constant molecular excitation temperature $T_{\rm ex}$ (corresponding to the rotational transition $J=3\rightarrow 2$) as the characteristic brightness temperature $T_{\rm b}$ associated with emission from all three species. We model the main beam temperature $T_{\rm MB}$, as
\begin{equation}\label{Tr}
  T_{\rm MB}=f[J_{\nu}(T_{\mathrm {ex}})-J_{\nu}(T_{\mathrm{bg}})](1-e^{-\tau_{\nu}}).
\end{equation}
%Here,  $T_{\text{bg}}$ is taken as the cosmic microwave background temperature ($\approx$ 2.73 K),
Here, $f$ is the beam-filling factor and is assumed to be 1. We assume the $^{12}$CO(3-2) line is optically thick, particularly near the line centre, so the excitation temperature can be approximated as 
\begin{equation}\label{Tex}
    T_{\text{ex}}=\frac{h\nu/k}{\ln\left[1+\frac{h\nu/k}{T_{\rm max}+J_\nu(T_{\rm bg})}\right]},
\end{equation}
where $T_{\rm max}$ is the peak of the $^{12}$CO(3-2) spectral distribution along each line of sight. Following \citet{mangum}, we have the column density of the top state of the transition for $^{13}$CO(3-2) and C$^{18}$O(3-2) expressed in terms of their optical depth integrated over the Doppler-broadened spectral profile for every position \citepalias[e.g.,][]{deb18}, 
\begin{equation}\label{Nu}
     N_u =\frac{8\pi \nu_0^3}{c^3 A_{ul}}\frac{1}{e^{\frac{h\nu_0}{kT_{\text{ex}}}}-1}\int \tau_\nu dv.
\end{equation}
Here, $\nu_0$ is the equivalent rest frequency and $A_{ul}$ is the Einstein coefficient for the $u=3$ to $l=2$ transition. We extrapolate total column density of the species using the partition function $Q$, which is well approximated as \begin{equation}
Q\approx \frac{kT}{hB_0}\exp\left(\frac{hB_0}{3kT}\right). 
\end{equation}
With these assumptions, the total column density is 
\begin{eqnarray}\label{Ntot}
    N_{\text{tot}}&=&\frac{Q}{g_u}\exp\left(\frac{E_u}{kT_{\text{ex}}}\right) N_u \nonumber \\
    &=& \frac{Q}{g_u}\exp\left(\frac{E_u}{kT_{\text{ex}}}\right) \frac{8\pi \nu_0^3}{c^3 A_{ul}}\frac{1}{e^{\frac{h\nu_0}{kT_{\text{ex}}}}-1}\int \tau_\nu dv.
\end{eqnarray}
For the C$^{18}$O line, the Einstein coefficient $A_{ul}=6.011\times 10^{-7}$ s$^{-1}$, $\nu_0=330.588$ GHz and the rotational constant $B_0=54891.42$ MHz. These values are obtained from LAMDA \footnote{\url{http://home.strw.leidenuniv.nl/~moldata/}} \citep{lamda} and NIST\footnote{\url{ https://physics.nist.gov/PhysRefData/MolSpec/}} databases.

In star-forming clouds, C$^{18}$O has a low abundance relative to $^{12}$CO ($N_\mathrm{C^{18}O}/N_\mathrm{{}^{12}CO}\approx 1.5 \times 10^{-3}$) and $^{13}$CO ($N_\mathrm{C^{18}O}/N_\mathrm{{}^{13}CO}\sim 0.1$) \citep{wilson94}, so it's often reasonable to assume the C$^{18}$O emission is optically thin. However, the line can be optically thick in some regions of star formation as some authors have suggested \citep{wh15}. In our case, we verify this by following the approach outlined in \citep{wh15, lad98} to estimate the line-of-sight maximum optical depth of C$^{18}$O (3-2) emission.  This approach compares the brightness ratio of $T_\mathrm{^{13}CO}/T_\mathrm{C^{18}O}$ to an assumed abundance ratio of $8$. Finding a brightness ratio significantly smaller than the abundance ratio would imply significant opacity in the C$^{18}$O line. We estimate the C$^{18}$O optical depth for the outflows for every pixel in the regions of significant emission. Then we compute medians of these values for each outflow. The median varies from 0.23 to 0.65 with corresponding standard deviation from 0.04 to 0.28. This may justify the treatment of C$^{18}$O (3-2) line as optically thin. In this case, the optical depths for the two species are derived from Equation \ref{Tr}, in terms of their main beam temperatures,
\begin{eqnarray} \label{tau}
     {\rm C^{18}O}: \tau_{\nu} &=& \frac{ T_\text{MB}}{J_\nu (T_{\rm ex})-J_\nu(T_{\text{bg}})}\\
    {\rm {}^{13}CO}:
    \tau_{\nu}&=& -{\rm ln}\left[1-\frac{T_{\rm MB}}{J(T_{\rm ex})-J(T_{\rm bg})}\right]
\end{eqnarray}

%%%%%%%%%%%%%%%%%%%%%%%%%%%%%%%%%%%%%%%%%%%%%%%%%%%%%%%%%%%%%%%%%%%%%%%%%%%%%%%%%%%%%%%%%%%%%%%%%%%%%%%%%%%%%%%%%%%%%%%%%%%%%%%%%%
\subsection{Physical properties of the outflows}\label{wings}
We estimate the mass, momentum, and kinetic energy of each outflow given the CO column density measured as a function of line-of-sight velocity. We use the $^{13}$CO(3-2) spectral line as the primary tracer of column density, since the optically thick low-$J$ transition lines of $^{12}$CO is subject to self absorption and will provide an underestimate of the mass near the line centre.
% overestimate the gas column density in outflow wings \citep{Dun}. 
The $^{13}$CO(3-2) line has low signal-to-noise ratio in outflow wings. At large velocity offsets, it can be too weak to extract any useful information. Hence, we implement an extrapolation technique, adopted from \citet{arce01}, for inferring $^{13}$CO(3-2) emission from the brighter $^{12}$CO(3-2) line. Using Equation \ref{Ntot}, we express the column density of $^{13}$CO(3-2) as a function of position offset (spatial pixel) and velocity along the spectral axis in a position-velocity (PV) slice \citepalias{deb18}, 

\begin{figure*}
    \centering
\includegraphics[width=\textwidth]{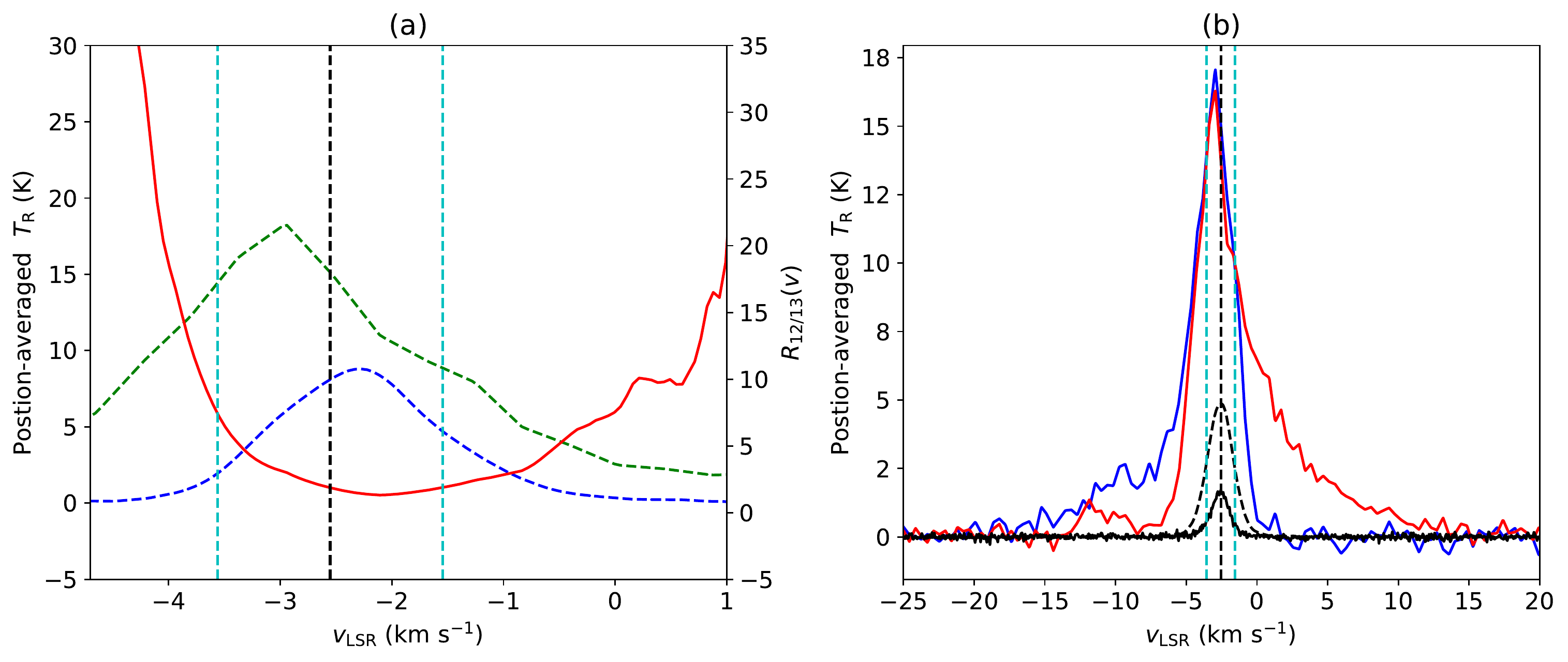}
    \caption{(a) Parabolic shape of $R_{12/13}(v)$ (red), plotted along with $^{12}$CO(3-2) (green),$^{13}$CO(3-2) (blue) emissions. Local minimum occurs near the emission peak.(b) Spectral line profiles of $^{12}$CO(3-2) (blue,red),$^{13}$CO(3-2) (dotted), and C$^{18}$O  (3-2)(solid) show relative brightness values around the line centre that is best identified by C$^{18}$O. Bipolar outflow is best visible in $^{12}$CO(3-2) line, where $^{13}$CO(3-2) emission is insignificant. Vertical dotted lines in both panels denote velocity centroid (black) and 4$\sigma$ limits (cyan) of the fitted Gaussian. Both diagrams are constructed from the data associated with outflow G81.435+2.147.\label{ratio}}
\end{figure*}

\begin{equation}\label{N13}
   N_{^{13}\rm CO} (x,y,v)=\frac{8\pi \nu_0^3Q_{\rm rot}}{7c^3 A_{ul}}\frac{e^{\frac{E_u}{kT_{\text{ex}}}}}{e^{\frac{h\nu_0}{kT_{\text{ex}}}}-1}\tau_\nu (\mathbf{x},v)\, \delta v.
\end{equation}
For calculating outflow properties we separate the asymmetric blue- and red-shifted wings of the spectral profile from the symmetric central components. As a first step, we estimate the velocity centroid $\varv_0$ of the line by fitting a Gaussian model to the C$^{18}$O(3-2) data, along with line width $\sigma_{\varv}$. We use C$^{18}$O(3-2) because it is optically thin and has a symmetric profile that is bright only around the line centre (Figure \ref{ratio}). We repeat this fitting process for every spatial pixel along the PV-slice (Figure \ref{pv}).  After fixing the line centre, we fit a quadratic function to the observed emission ratio of $^{12}$CO(3-2)/$^{13}$CO(3-2), denoted by $R_{12/13}(v)$ since the line ratio typically resembles a parabola around $\varv_0$:
$$R_{12/13}(v) \hspace{1mm} \hat{=} \hspace{1mm} C_0 + C_2\hspace{0.5mm}(\varv - \varv_0)^2.$$
This is done separately for each of the blue- and red-shifted lobes. We also set an upper limit of 65 for the ratio, based on the relative abundance of the two molecular species in molecular clouds \citep{wilson94}. The fitted ratio ranges between this value and a minimum at the velocities where the $^{13}$CO(3-2) line is the brightest (Figure \ref{ratio}). Using the main beam temperature of $^{12}$CO(3-2) ($T_{12}$) and the fitted ratio $R_{12/13}$, we can infer the $^{13}$CO(3-2) main beam temperature ($\hat{T}_{13}$) where the signal is undetectable.  Following a strategy adapted from \citet{arce01}, we estimate $\hat{T}_{\rm R,13}(x,y,v)$ in three regimes based on the signal-to-noise ratio of the two emission lines:
\[\hat{T}_{13}(x,y,v)=
    \begin{cases} 
          T_{13} & \mathrm{if} \hspace{2mm} T_{13} \geqslant 5 \hspace{1 mm} \sigma_{13}\\
         \frac{T_{12}}{R_{12/13}} & \mathrm{if} \hspace{2mm} T_{13} < 2 \hspace{1 mm}\sigma_{13},\hspace{1 mm} T_{12} \geqslant 2 \hspace{1 mm}\sigma_{12} \\
      0 & \mathrm{if} \hspace{2mm} T_{13}  < 2 \hspace{1 mm} \sigma_{13},\hspace{1 mm}T_{12}<2 \hspace{1 mm}\sigma_{12}. 
   \end{cases}
\]
Here, the noise levels of the two lines are given as $\sigma_{12}$ and $\sigma_{13}$. The last condition states that $^{13}$CO(3-2) main beam temperature cannot be estimated when both emission lines are undetectable.

% This is possible since $^{12}$CO(3-2) is optically thick at all velocities where $^{13}$CO(3-2) signal is above its root-mean-square noise level, denoted by $\sigma_{13}$.

Using the optical depth (Equation \ref{tau}) and the column density (Equation \ref{N13}), we determine the H$_2$ column density $N_{\rm H_2}(x,y,\varv)$ as $N_{\rm H_2}=N_{^{13}\rm CO}/X_{\mathrm{CO}}$ by assuming a fixed abundance ratio $X_{\mathrm{CO}}=10^{-6}$ of $^{13}$CO(3-2) relative to H$_2$ \citep{wilson94}. For the total mass in the outflow, we integrate $N_{\rm H_2}(x,y,v)$ over blue- and red-shifted segments of the spectral axis at each position along the PV-slice and then sum over all such positions,
\begin{equation}
\begin{split}
    M_{\rm outflow} &=\mu_{\rm H_2} \int_{\mathbf{x}} \int_{\rm wing} N_{\rm H_2}(x,y,\varv)\, d\nu \, d\mathbf{x} \\
    &\approx \mu_{\rm H_2} \sum_{\mathbf{x}, \rm wing} N_{\rm H_2}(x,y,\varv)\, \delta \varv\, \delta A_{\rm pix}.  
\end{split}
\end{equation}
Here, we have assumed a mean molecular mass per H$_2$ as $\mu_{\rm H_2} = $2.4 $m_{\rm H}$ ($m_{\rm H}$ assuming a 10\% atomic He abundance by number.  We determine the physical pixel areas by projecting the angular size of each pixel to the assumed distance of the outflow (Section \ref{sec:distances}; Table \ref{prop_compare}).

% The distances of the outflows from the sun (see table \ref{prop_compare}) were used for calculating pixel area. 

We also estimate the projected outflow momentum and energy using 
\begin{eqnarray}
     p = p_0 \cos \theta & =&\sum M(x,y,\varv) |\varv-\varv_0|\\ E=E_0 \cos^2 \theta & = &\frac{1}{2}\sum M(x,y,\varv) (\varv-\varv_0)^2
\end{eqnarray}
where $\theta$ is the unknown inclination angle with respect to the line of sight and $p_0$ and $E_0$ indicated the unprojected momentum and energy. The results are summarized in Table \ref{prop_compare}.  
%It is to be noted that the inclination angles have scope for large uncertainties so these outflow parameters are likely underestimated and can be up to two orders of magnitudes larger than the estimated values \citep{Dun}.  
Table \ref{proto_association} includes the protostellar sources that generate the outflows, identified by searching the \citet{kry} catalogue, along with their infrared (IR) luminosity, except the source NOMAD1 1323-0477179 for which we were unable to find the spectral index value and IR luminosity.

We exclude the source G80.314+1.330 from further analysis. We were unable to find a protostar associated with the object G80.314+1.330 in existing catalogues. \citet{gott12} identified this object as an outflow, which we also have confirmed using spectral distribution, contour, and PV plots (Figure \ref{of7}). However, the high negative velocity (Table \ref{obs_compare}) and weak emission suggest that this outflow is unlikely to be located in Cygnus X but is likely located further away along the line of sight, likely in the Perseus arm.  Hence, a protostar is too distant to be detected.

%\begin{sidewaystable}
%\begin{landscape}
\begin{table*}
%\Rotatebox{90}{%
%\pagestyle{empty}
%\hfill
%\begin{adjustbox}{angle=90}  
    \centering
\captionof{table}{Dynamical properties of the 12 outflows: mass, momentum, and energy columns for estimates from all three CO lines, discussed in section \ref{wings}, along with corresponding estimates from $^{12}$CO alone, discussed in section \ref{12to13}.} 
\label{prop_compare}
    \begin{tabular}{cccccccccc}
    \hline
    Outflow & Distance  & $T_{\rm ex}$ & Mass & $^{12}$CO-only Mass& Momentum& $^{12}$CO-only Momentum & Energy &$^{12}$CO-only Energy   \\
    &(kpc) & (K) &(M$_\odot$)&(M$_\odot$)&(M$_\odot$) km s$^{-1}$ &(M$_\odot$) km s$^{-1}$ &($10^{45}$ergs)&($10^{45}$ergs) \\
         \hline
         \vspace{2mm}
           G79.886+2.552 &  $0.65\pm 0.15$ & 16 & $0.72\pm0.16$ & $0.42\pm0.33$ & $3.86\pm0.89$ & $1.82\pm1.10$ & $0.34\pm 0.08$ &  $0.15\pm0.10$\\
           \vspace{2mm}
           G81.435+2.147& $1.4\pm 0.08$& 25& $5.12\pm 0.29$ & $1.42\pm0.63$ & $27.01\pm1.55$ & $6.40\pm1.27$ & $1.95\pm0.11$ & $0.40\pm0.23$\\
          \vspace{2mm}
           G81.424+2.140& $1.4\pm0.08$ & 22 & $1.36 \pm0.06$ & $0.70\pm0.31$ & $5.70\pm0.29$ & $2.31\pm0.40$ & $0.30\pm 0.02$ & $0.09\pm0.05$\\
          \vspace{2mm}
           G81.302+1.055& $1.3\pm0.07$ & 36 & $9.68\pm0.52$ & $5.18\pm0.59$ & $30.20\pm1.62$ & $18.72\pm2.06$ & $1.15\pm0.06$ & $0.97\pm0.11$\\
          \vspace{2mm}
           G80.862+0.385& $1.5\pm 0.08$ & 31 & $4.65\pm0.24$ & $4.70\pm0.54$ & $31.8\pm1.66$ & $22.31\pm2.51$ & $1.48\pm0.11$ & $1.34\pm0.13$\\
         \vspace{2mm}
          G81.663+0.468& $1.3\pm 0.07$ & 19 & $2.90\pm0.17$ & $1.56\pm0.14$ & $21.30\pm1.15$ & $14.58\pm1.30$ & $2.06\pm0.16$ & $1.50\pm0.14$\\
          \vspace{2mm}
          G81.551+0.098& $1.5\pm 0.08$ & 17 & $1.43\pm0.08$ & $0.33\pm0.04$ & $2.94\pm0.16$ & $0.80\pm0.09$ & $0.08\pm0.004$ & $0.05\pm0.002$\\
          \vspace{2mm}
          G81.582+0.104&  $1.5\pm 0.08$ & 24 & $3.68\pm0.20$ & $1.41\pm0.15$ & $7.50\pm0.40$ & $8.00\pm0.85$ & $1.69\pm0.01$ & $0.47\pm0.05$\\
          \vspace{2mm}
           G82.581+0.203& $1.3\pm 0.07$ & 20 & $2.13\pm0.11$ & $1.27\pm0.14$ & $12.84\pm0.68$ & $8.21\pm0.94$ & $1.48\pm0.08$ & $0.74\pm0.08$ \\
           \vspace{2mm}
           G82.571+0.194&$1.3\pm 0.07$ & 18 & $1.16\pm0.06$ & $0.33\pm0.04$ & $3.38\pm0.18$ & $1.25\pm0.14$ & $0.22\pm0.01 $ & $0.08\pm0.01$ \\
          \vspace{2mm}
           G80.158+2.727 & $0.65\pm 0.07$ & 16 & $1.50\pm0.16$ & $0.32\pm0.07$ & $6.22\pm0.68$ & $1.21\pm0.23$ & $0.43\pm0.05$ & $0.06\pm0.01$\\
           \vspace{2mm}
           G80.149+2.710 & $0.65\pm 0.07$ & 27 & $0.18\pm0.02$ & $0.36\pm0.08$ & $1.13\pm0.12$ & $1.10\pm0.23$ & $0.13\pm0.01$ & $0.08\pm0.01$\\
         \hline
    \end{tabular}
\end{table*}
%\end{landscape}
%\end{sidewaystable*}
%\end{adjustbox}
%\hfill
%\null
%}%
%\end{sidewaystable}  

\begin{table*}
\centering
\captionof{table}{Protostellar sources associated with the 12 outflows, as identified in \citet{kry}.}
\label{proto_association}
    \begin{tabular}{cccccc}
    \hline
    Outflow & Distance & IR Source & Angular & Spectral & $L_{\mathrm{IR}}$  \\
      & (kpc) & & Separation & Index & log(L/L$_\odot$) \\
         \hline
         \vspace{2mm}
           G79.886+2.552 & $0.65\pm 0.15$ & J202430.49+420409.19 & $16.42^{''}$ & 0.16 & 1.87\\
           \vspace{2mm}
           G81.435+2.147& $1.4\pm 0.1$&  J203111.82+430521.66 &$21.70^{''}$&2.12 &0.84  \\
          \vspace{2mm}
           G81.424+2.140 & $1.4\pm 0.1$& J203112.70+430457.56 &$6.30^{''}$&0.91 &0.45 \\
          \vspace{2mm}
           G81.302+1.055 & $1.3\pm 0.1$ & J203534.44+422006.80 &$14.58^{''}$&1.23 &1.95 \\
          \vspace{2mm}
           G80.862+0.385& $1.5\pm 0.1$& J203702.60+413440.97 &$8.76^{''}$&1.34 &1.72\\
         \vspace{2mm}
          G81.663+0.468& $1.3\pm 0.1$ &J20391672+4216090.00 &$10.94^{''}$&$-0.05$ &2.39 \\
          \vspace{2mm}
          G81.551+0.098& $1.5\pm 0.1$ & J204028.48+415711.97 &$9.36^{''}$&1.64 &2.04 \\
          \vspace{2mm}
          G81.582+0.104&  $1.5\pm 0.1$ & J204033.48+415900.63&$4.81^{''}$&1.87 &0.95  \\
          \vspace{2mm}
           G82.581+0.203& $1.3\pm 0.1$ & J204322.87+425022.76 &59.64$^{''}$& 0.93 & $-0.58$ \\
           \vspace{2mm}
           G82.571+0.194&$1.3\pm 0.1$ & J204328.27+424900.09 &$11.64^{''}$& 0.82&2.28 \\
          \vspace{2mm}
           G80.158+2.727 & $0.65\pm 0.15$ & J202434.18+422331.60 & $19.28^{''}$&0.84 &1.56\\
           \vspace{2mm}
           G80.149+2.710 & $0.65\pm 0.15$ & NOMAD1 1323-0477179&17.51$^{''}$& $\cdots$ & $\cdots$ \\
         \hline
    \end{tabular}
\end{table*}
%\clearpage
\medskip

%%%%%%%%%%%%%%%%%%%%%%%%%%%%%%%%%%%%%%%%%%%%%%%%%%%%%%%%%%%%%%%%%%%%%%%%%%%%%%%%%%%%%%%%%%%%%%%%%%%%%%%%%%%%%%%%%%%%%%%%
\section{Estimation of outflow properties based on $^{12}$CO(3-2) data}\label{12to13}
%For a

Outflows are ubiquitous in wide-area surveys of molecular emission \citep{gott12,drabek16}, and the feedback from outflows into the molecular ISM is best understood in the context of these large surveys. However, a full determination of outflow properties requires multiple isotopologues (Section \ref{allco}) and, ideally, multiple rotational transitions from those isotopologues to measure both opacity and excitation temperature \citep{Dun}. While ideal, observing all these transitions is expensive in terms of telescope time, so approximate methods are needed to interpret survey data.

% Many class 0 and I young stellar objects embedded in dust envelopes exhibit bipolar outflows with detectable high-velocity $^{12}$CO(3-2) line emission, but the estimation of gas column density is less reliable using optically thick $^{12}$CO(3-2) line. 

% As \cite{Dun} mentioned proper multi-tracer observation of outflows involving non-optically thick $^{13}$CO lines, which provide much better estimates of column densities are 
% expensive in terms of telescope time, a suggested use of an optical depth correction factor $\frac{\tau_{12}}{1-e^{-\tau_{12}}}$ is found in the literature for estimating mass from $^{12}$CO(3-2) line which is bright in the outflow wings and excellent for detecting outflows \citep{Zh20,Plun, Dun, Gins11}. However, even with the correction factor the mass estimate from  $^{12}$CO(3-2) alone would still be an underestimation by several factors to an order of magnitude \citep{Gins11}, possibly because the assumption of $^{13}$CO(3-2) to be optically thin in the outflow wings may not be valid for lower velocity offsets from the line center. 

To analyze outflows in the wide area survey of Cygnus X (Deb et al. in preparation), we need to estimate outflow mass and other dynamical properties without $^{13}$CO(3-2). A common approximation is to estimate an optical depth correction factor, $\frac{\tau_{12}}{1-e^{-\tau_{12}}}$, to measure out mass from the $^{12}$CO(3-2) line alone \citep{Zh20,Plun, Dun, Gins11}. However, even with the correction factor the mass estimate from  $^{12}$CO(3-2) alone could still be an underestimate by 0.5 to 1 dex \citep{Gins11},  because the assumption of $^{13}$CO(3-2) to be optically thin in the outflow wings may not be valid for lower velocity offsets from the line centre (refer to section \ref{upe}). 

Here, we use our in-hand data on $^{13}$CO emission to calibrate empirical relationships between the observed $^{12}$CO(3-2) emission ($T_{12}$) and the outflow properties as characterized from the full analysis of the $^{13}$CO(3-2) data (section \ref{allco}). Specifically, we empirically estimate the opacity that would be seen in the $^{13}\mathrm{CO}$ line, which we infer based on the brightness of the $^{12}\mathrm{CO}$ emission. The empirical estimate avoids using the (unobserved) $T_\mathrm{MB}$ for $^{13}$CO and scales the $^{12}$CO brightness directly to the $^{13}$CO optical depth. We also estimate the line centroid and width so we can define the velocity ranges that correspond to the wings of the outflow and the velocities relative to the line centre.

\begin{figure}
               \includegraphics[width=\columnwidth]{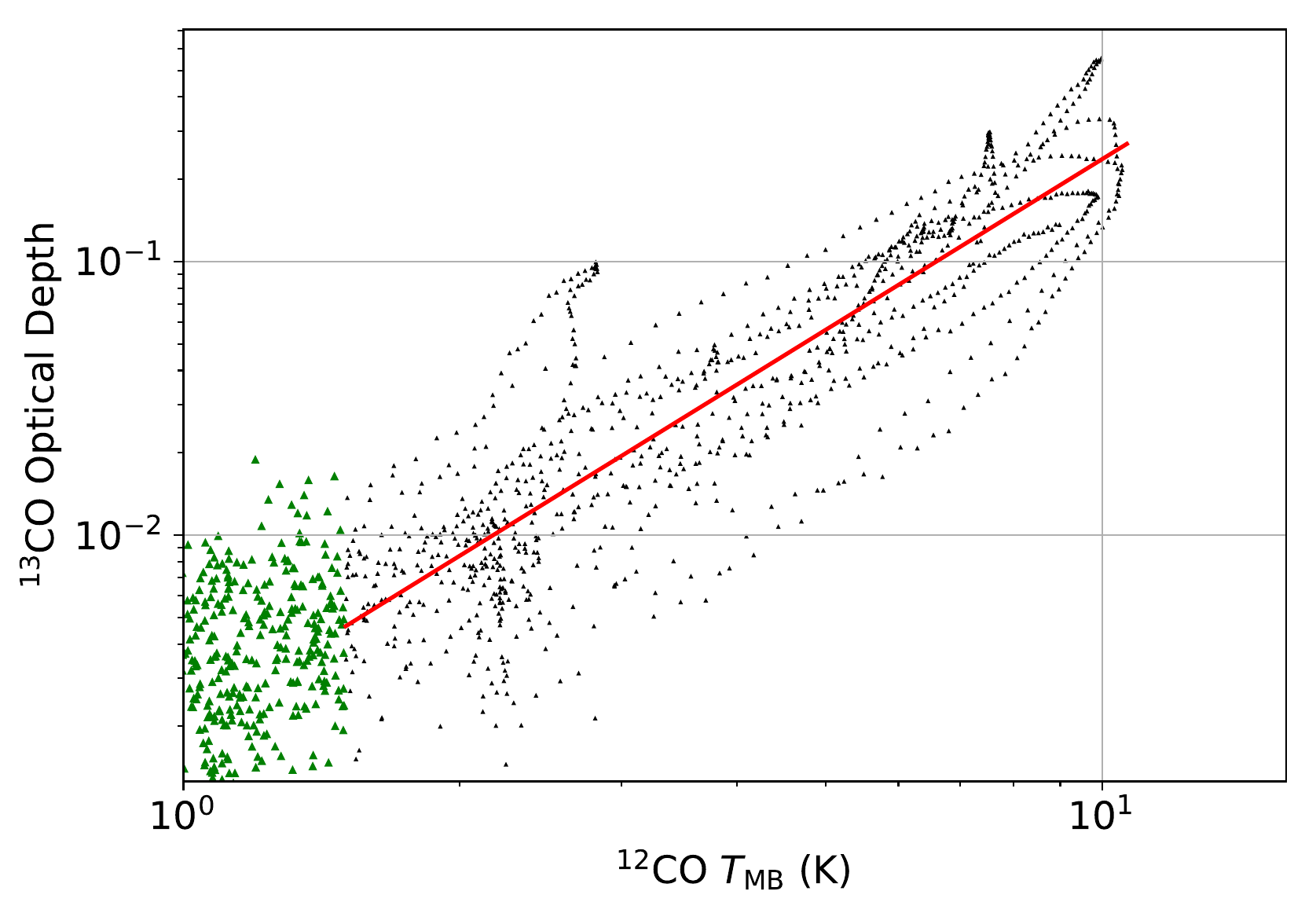}
               \caption{Scatter plot shows the association between $^{12}$CO(3-2) emission in terms of position-averaged main beam temperature in K and $^{13}$CO(3-2) optical depth. The raw data set is divided into detectable signal (in black) and noise ($<2\sigma_{12}$, in green). The straight line (in red) denotes the line of best fit. \label{curve_fit}}
\end{figure} 

Our empirical relationship between $^{12}\mathrm{CO}$ brightness and $^{13}\mathrm{CO}$ optical depth is shown in Figure \ref{curve_fit}, where we fit a linear relationship between the log of both quantities.  Since the $^{12}$CO(3-2) line observations were stored as a data cube (Figure \ref{pv}), we average the value $T_{12}(x,y,\varv)$ over the position coordinate. Similarly, our estimate of $\tau_{\nu,13}$ is from the full analysis in \ref{allco}, and we again average$\tau_{\nu,13} (x,y,v)$ over position coordinates. Figure \ref{curve_fit} shows the scatter plot of $\left(\tau_{\nu,13}, T_{12}\right)$, for all outflows included in Table \ref{prop_compare}. We perform a linear regression on the bivariate set, with adjusted-R$^2=0.8$ and F-statistic$= 3538$ demonstrating a strong relationship.  The best fit in log-space is given by, 
\begin{equation}\label{logmodel}
    {\rm log}_{10} \tau_{\nu,13} = -2.69\pm 0.02 + (2.07\pm 0.04) \times {\rm log}_{10} T_{12} .
\end{equation}

\begin{figure}
               \includegraphics[width=0.9 \columnwidth]{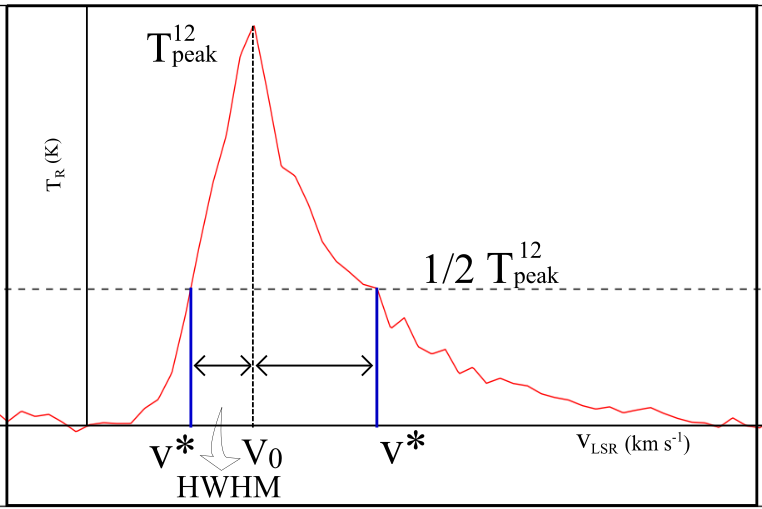}
               \caption{A schematic view of HWHM estimation technique from a spectral line profile \label{hwhm}}
\end{figure} 

Using this fitted equation, we estimate $^{13}$CO column density as a function of position and velocity in a PV-slice, again by assuming a mean particle mass of $\mu_{\rm H_2}=2.4 m_{\rm H}$ and using distances of the outflows from the sun.

To estimate the wing mass, we also estimate the profile line centre $\varv_0$ and velocity width $\sigma_\varv$. Unlike in Section \ref{wings}, here we assume we do not have access to $^{13}$CO(3-2) and C$^{18}$O(3-2) data, so we approximate the $^{12}$CO(3-2) spectral line with a Gaussian profile. We then estimate the line centre by leaving $\varv_0$ as a free parameter and minimizing the outflow kinetic energy along each line of sight in the PV slice. 

% at each position along the PV-slice and then estimate the centroid and line width by comparing the assumed properites to the .
%i.e., $$ \frac{\partial}{\partial \varv_0}\sum_\varv M(x,y,\varv)  \left(\varv-\varv_0\right)^2|_{\varv_0=\mu_\varv}= 0$$.  Here $\varv - \varv_0$ denotes velocity offset. 

Next, we calculate $\sigma_\varv$ by measuring the half width at half maximum (HWHM) of the line profile, where for a Gaussian, $\mathrm{HWMW}= \sqrt{2 \hspace{1mm}{\rm ln}\hspace{1mm} 2}\,\sigma_\varv$. Since the line profile is asymmetric, we measure the HWHM on both sides of $\varv_0$ and take the minimum width as the line width as shown in the schematic Figure \ref{hwhm}.  We measure the HWHM by finding the velocity channels $\varv^*$ corresponding to the brightness $\frac{1}{2}T_{\rm peak}^{12}$ where $T_{\rm peak}^{12}$ denotes the maximum of $T_{\rm MB}$ for a spectral profile. In that case, referring to figure \ref{hwhm}, we can write, \[ \hspace{3cm}{\rm HWHM}=\min_{\varv^{*}}  |\varv^{*} - \varv_0|.\]
There is foreground absorption observed in the outflow spectra (Figure \ref{of1}), which is possibly caused by the foreground Cygnus Rift. This absorption feature, however, does not alter the estimation of $\varv_0$ and $\sigma_\varv$ because the outflow wings are unaffected by the absorption. The inferred value of $\sigma_\varv$ can be up to a factor of two larger than the value measured directly from the observed $^{13}$CO(3-2) line.  
%, resulted from the fact the inferred $^{13}$CO(3-2) emission exists at large velocity offsets from the line center where the observed emission falls below the detection threshold. 

\begin{figure*}
     \centering
     \includegraphics[width=\textwidth]{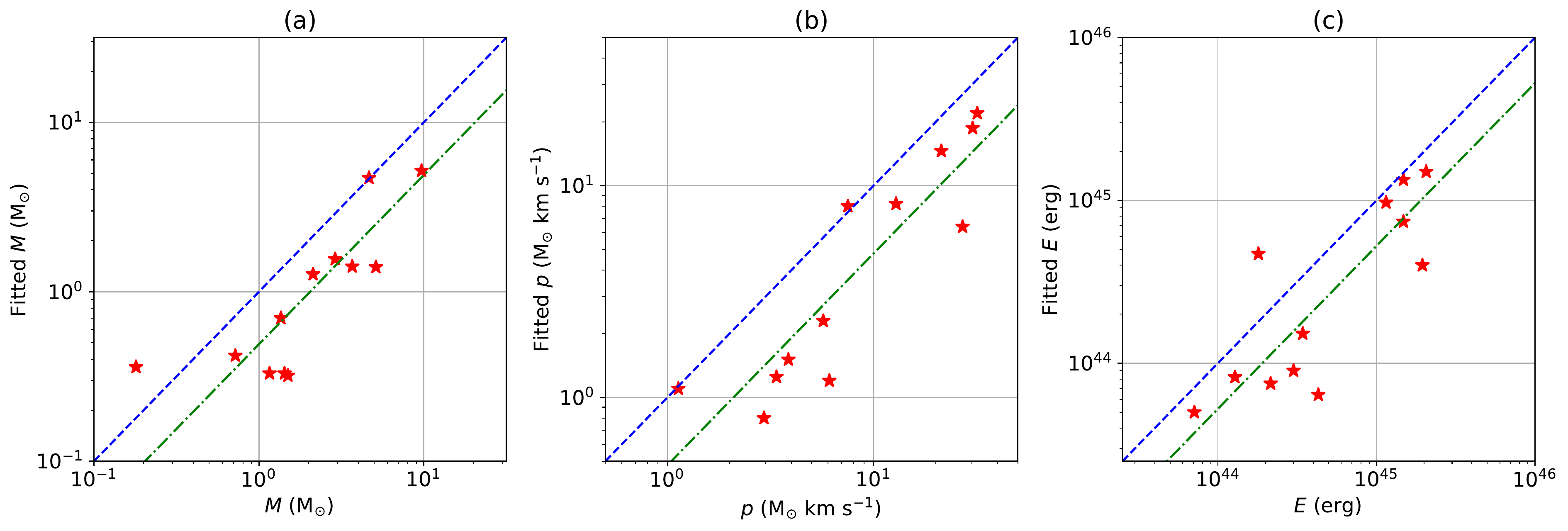}
     \caption{Scatter plots show comparison between outflow mass ($M$), projected momentum ($p$) and projected energy ($E$) estimated from $^{13}$CO(3-2), $^{12}$CO(3-2) and C$^{18}$O(3-2) data (x-axis) and those estimated from $^{12}$CO(3-2) alone (y-axis). Blue dashed lines denote perfect correlation. Green dash-dotted lines denote the relationship between three-line estimated values $^{12}$CO-only values. A comparison between the two sets of lines shows a consistent underestimation of the outflow properties. \label{corr}}
   \end{figure*}

We define the outflow velocity wings as spectral regions with $|\varv - \varv_0| > 2\sigma_\varv$. The mass estimate is obtained by summing over such velocity channels and position offsets along the PV-slice. The estimated values of mass, projected momentum and projected energy are included in Table \ref{prop_compare}. Similar to section \ref{wings}, momentum and energy values estimated from $^{12}$CO alone contain unknown projection angle with respect to the line of sight. In Figure \ref{corr}, we compare the property estimates for the $^{12}\mathrm{CO}$-only method vs those derived from using all three lines. Considering the small sample size, there is good correlation between the fitted the estimated values but some measurable systematic differences. Table \ref{12-13} summarizes the typical differences.  The mean mass from $^{12}$CO alone is typically 0.31 dex (a factor of 0.48) smaller than the estimates from all CO lines.  The momentum and energy values are also a factor of 0.47 and 0.53 smaller than the corresponding estimates from all CO lines. The consistent slight underestimation of outflow energetics is attributed to the larger inferred $^{12}$CO(3-2) value of $\sigma_\varv$ mentioned before. Table \ref{12-13} also notes the width of the distribution, which is comparable to the offset that we measure.  We do not apply any ad hoc scalings at this point to the $^{12}\mathrm{CO}$-only estimates to bring them into agreement with the full analysis, but we will consider the offsets and spread in Table \ref{12-13} as part of our error budget.

\begin{table}
    \begin{tabular}{cccl}
    \hline
    $\log_{10}$($^{12}$CO-only / All lines) quantities & Mean & Standard deviation \\
        \hline
         \vspace{1mm}
          Mass& $-0.31$ & 0.26\\
          \vspace{1mm}
           Momentum & $-0.32$ & 0.23\\
          \vspace{1mm}
          Energy & $-0.28$ & 0.31\\
         \hline
    \end{tabular}
    \caption{Comparison between outflow properties from the approximations using the $^{12}$CO(3-2) line alone and those estimated from all three CO lines. On average, this approach systematically underestimates dynamical properties by $\sim 0.3$~dex, which should be included in an error budget.}
\label{12-13}
\end{table}

%%%%%%%%%%%%%%%%%%%%%%%%%%%%%%%%%%%%%%%%%%EXTRA%%%%%%%%%%%%%%%%%%%%%%%%%%%%%%%%%%%%%%%%%%%%%%%%%%%%%%%%%%%%%%%%%

 %%%%%%%%%%%%%%%%%%%%%%%%%%%%%%%%%%%%%%%%%%%%END%%%%%%%%%%%%%%%%%%%%%%%%%%%%%%%%%%%%%%%%%%%%%%%%%%%%%%%%%%%%%%%%
     \begin{figure*}
     \centering
     \includegraphics[width=\textwidth]{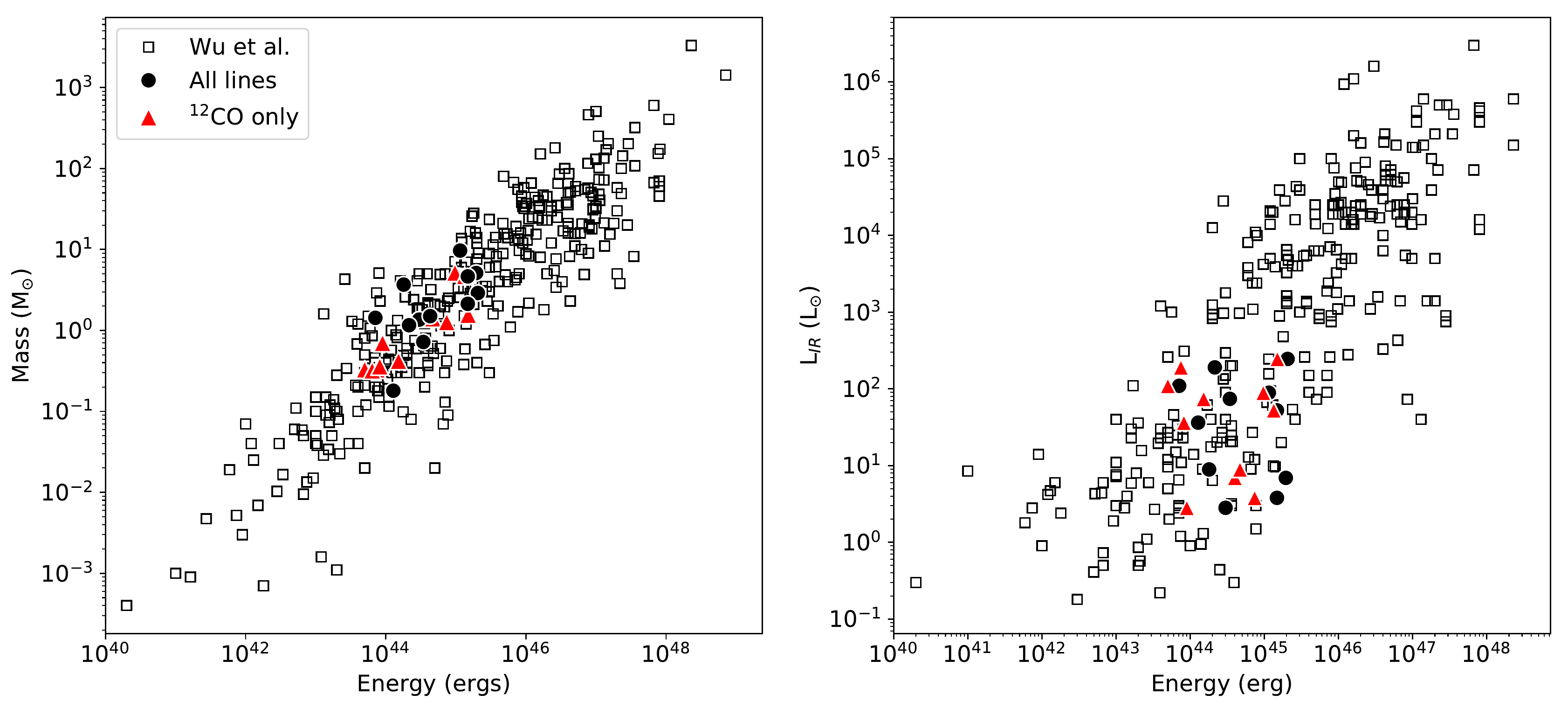}
     \caption{(left) Scatter plots show outflow mass plotted against energy. Red triangles denote quantities estimated from only $^{12}$CO(3-2) data. (right) Infrared luminosity plotted against outflow energy. Colour scheme is the same as in (left).\label{wu} The Cygnus X outflows are consistent with the broader population irrespectrive of the method used for property estimation.}
   \end{figure*}

%%%%%%%%%%%%%%%%%%%%%%%%%%%%%%%%%%%%%%%%%%%%%%%%%%%%%%%%%%%%%%%%%%%%%%%%%%%%%%%%%%%%%%%%%%%
\section{Discussion} \label{discuss}

\subsection{Outflow Properties and Protostellar Sources}
We have estimated several dynamical properties of 12 outflows and identified their infrared sources using the \citet{kry} catalogue. Based on the spectral index ($\alpha$) value, we categorize J202430.49+420409.19 and J20391672+4216090.00 as flat-spectrum protostars. All of the remaining protostars have spectral index  values (defined as $\alpha$ in $F_\nu \propto \nu^{-\alpha}$) greater than 0.3. These values imply they are in their early stages of evolution and belong to either class 0 or class I.  The early evolutionary stages also implies the bolometric luminosity is approximately the same as infrared luminosity ($L_{\rm IR}$), included in Table \ref{prop_compare}. We were unable to find luminosity and spectral index value for NOMAD1 1323-0477179, which was referred as the IR source associated with the outflow G80.149+2.710 in \citet{gott12}. It is likely that the IR source of this outflow is a deeply embedded class 0 protostar in its early stages of evolution.   

As suggested from previous analyses \citep[][and references therein]{bally16}, we examined the correlation between mechanical luminosity $L_{\rm mech}$, or infrared luminosity $L_{\rm IR}$ and spectral index. As defined, a lower value of spectral index indicates a more evolved protostar. Outflow energy, IR and mechanical luminosities nominally decrease as the protostar evolves, the highest being achieved in the early stages, we would expect a positive correlation of these outflow properties with increasing spectral index. In our sample such correlation is observed but is extremely weak with large scatter.  We attribute this to having a small, heterogeneous sample of outflows at various distances and the narrow range of spectral index that is recovered. We will revist these scalings in the context of the larger outflow survey (Deb et al., in preparation).  

For context, we compare our sample with the catalogue of \citet{wu}, which assembles a meta-analysis of outflow properties from the literature. The 12 outflows are broadly consistent with the population of outflows with respect to all their measured properties. In particular, we find that the mechanical luminosity $L_{\rm mech}$ is, on average $\sim 10^{-3} L_{\rm IR}$, which traces the accretion power, consistent with other sources.

%Additionally, we compare outflows G81.302+1.055 and G81.663+0.468 that are nearby, belong to different spectral classes and at the same distance of 1.3 kpc. Although G81.302+1.055 is more massive, spectral index implies G81.663+0.468 is more evolved to a different class (flat-spectrum) and is more energetic and luminous despite being less massive (Table \ref{prop_compare}).  

\subsection{Uncertainties in Parameter Estimates}\label{upe}

We have used CO lines for estimating outflow mass, momentum and energy, which are subject to significant uncertainties based on our assumptions. Even so, CO molecules remain the best species for studying the molecular outflows because of their high line intensity, low critical density, near-LTE excitation modes, and their relatively large abundances relative to other molecules.

Our estimates of outflow properties from a single $^{12}$CO line is similar to other approaches forwarded in the literature. Among early work involving CO lines, \citet{Bon} estimated outflow momentum flux from $^{12}$CO~(2-1) emission using $p\propto \int_{\mathrm{wings}} T_{\mathrm R}^{12}(\varv) d\varv A(r,dr) v^2$, $r$ denotes the radius of a projected annulus orthogonal to the outflow direction and $dr$ is the width of such annulus.  This is comparable to the approach discussed here, with modifications since the authors estimated momentum from $^{12}$CO emission in terms of radiation temperature and integrating over the spectral and spatial spread of the outflowing gas.  Another common assumption found in the literature is that outflow wings are optically thick in the $^{12}$CO line \citep{BL2, arce01, Dun,rohlfs}. An optically thick tracer only reflects the conditions of the surface of the cloud, thus results in an underestimation of mass, and subsequently of momentum and energy. We use the optically thinner $^{13}$CO(3-2) line for tracing H$_2$ column density in the outflow wings, although we have not made any explicit assumption that $\tau_{13}\ll 1$. Instead, we rely on the assumption of a constant excitation temperature for all lines and for all species. \cite{BL2} suggested a similar method for estimating wing column density from $^{13}$CO~(1-0) line. The authors used the observed $^{13}$CO emission when it was above the RMS noise level and extrapolated from $^{12}$CO~(1-0) using a second-order fitted polynomial ratio $R_{12/13}$ when $^{13}$CO was below the noise level. However, the authors used a different intrinsic abundance ratio, which provides a corresponding limit for the fitted brightness ratios ($R_{12/13} \leqslant 89$).
%Also, they differ in the assumption of optically thin high velocity $^{13}$CO lobes. \citet{rohlfs} argued such assumption would lead to overestimation of H$_2$ column density. 

Some authors suggested estimating mass from $^{12}$CO(3-2) brightness by using an opacity correction factor $\frac{\tau_{12}}{1-e^{-\tau_{12}}}$ \citep{Dun}. This is done by assuming $^{13}$CO(3-2) is optically thin, thereby numerically solving $\tau_{12}$ from the observed ratio $R_{12/13}$ using Equation \ref{Tr} under LTE, and here using a $^{12}$CO/$^{13}$CO abundance ratio of 65 \citep{wilson94},  
\begin{equation}\label{rat}
    R_{12/13}=\frac{T_{12}}{T_{13}} = \frac{1-e^{-\tau_{12}}}{1-e^{-\tau_{13}}}\approx 65 \frac{1-e^{-\tau_{12}}}{\tau_{12}} 
\end{equation}
$\frac{\tau_{12}}{1-e^{-\tau_{12}}}$ compensates for $^{12}$CO(3-2) being optically thick in line wings,  
\begin{equation}\label{T12}
    \hat{T}_{12} \sim 65\hspace{1mm} {T_{13}} = \frac{\tau_{12}}{1-e^{-\tau_{12}}}{T_{12}}. 
\end{equation}
The factor $\frac{\tau_{12}}{1-e^{-\tau_{12}}} \sim \tau_{12}$. \cite{rohlfs} note that Equation \ref{T12} would overestimate the ratio $R_{12/13}$ in Equation \ref{rat} by an amount that scales with $\tau_{12}$, resulting in an underestimation of the opacity correction factor. This underestimate arises because the assumption that $^{13}$CO(3-2) is optically thin may not be true near the line centre. The opacity profile can vary from one outflow to another. This ambiguity motivated our empirical model for determining the gas column density in outflow wings using the conditional estimation technique described in section \ref{wings}.

Our primary assumption is that all three CO lines are in LTE and have the same thermal excitation temperature $T_{\rm ex}$. \cite{Gins11} caution that while lower-$J$ transition lines of $^{12}$CO might be in LTE,  $^{12}$CO(3-2) may not be in LTE because of its high critical density value ($27\times$ greater than $J=1\rightarrow0$ line).  In this case, the $^{12}$CO(3-2) line may be subthermally excited ($T_{\rm ex}<T_{\rm K}$), which following the expression for $^{13}$CO(3-2) optical depth and Equation \ref{N13}, implies the gas column density is underestimated.  The mass and other dynamical properties would also then be lower limits. However, the authors mentioned their sample sizes were small and their claim of $^{12}$CO(3-2) being a poor tracer for column density in outflows is more relevant for later stages of evolution with warmer gas. By comparing the rotational transition lines of $^{12}$CO  \citet{Gins11} showed that the $J=3\rightarrow2$ line produces lower estimates of column density than the $J=2\rightarrow1$ and $J=1\rightarrow0$ lines for gas at higher excitation temperatures ($T_{\rm ex}>20$K). In contrast, \citet{Plun} measured mass and other dynamical properties by adopting specific fixed values of $T_{\rm ex}$ as well as a functional form of $T_{\rm ex}$ that varied from pixel to pixel. This may be generally better than our method of estimating column density $T_{\rm peak}$ of $^{12}$CO(3-2) emission in LTE as described in section \ref{wings}. However, the pixel-by-pixel $T_{\rm ex}$ profile does not produce significantly different values, unless the gas is warm ($T_{\rm ex}>$50 K) \citep{Plun}. In our case, the estimated excitation temperature ranges from 16 K to 36 K which, following the argument of \cite{Plun}, should produce results in good agreement with that from a more generalized temperature profile. 

We also developed a model for extrapolating H$_2$ column density from the $^{12}$CO(3-2) line alone. \cite{BL} estimated outflow mass of 12 sources from $^{12}$CO(3-2) and $^{13}$CO(3-2) lines by determining an assumed common excitation temperature by imposing a different 12-to-13 CO abundance ratio. The authors used a functional dependence of column density $N_{13}(\nu)$ on $T_{\mathrm ex}$ and $\tau_{13}$. For the sources with missing $^{13}$CO(3-2) data, they constrained $\tau_{12}\ll 1$ and $T_{\mathrm ex}>10$ K to estimate $^{12}$CO(3-2) column density, and used a fixed $^{12}$CO(3-2) to H$_2$ ratio. 
In contrast, we have not imposed restrictions on $\tau_{12}$ and $T_{\mathrm ex}$ for measuring H$_2$ column density. Instead, we used a direct approach of least square fitting to establish a functional relation between $\tau_{13}$ and $^{12}$CO(3-2) brightness. Since the two CO line species have approximately the same abundance ratio in all star-forming clouds, and $^{12}$CO(3-2) transition is ubiquitous in outflows of class 0 and I protostars, the advantage of our approach is that Equation \ref{logmodel} may be applicable in any outflow study that lacks $^{13}$CO(3-2) line data. 

This approach establishes a direct relationship between the two CO lines in the outflow wings with more generality. Figure \ref{corr} and Table \ref{12-13} summarize the small sample correlation between the fitted $^{13}$CO model and estimates based on all lines. There is a systematic underestimate of outflow properties, which may be caused by unaccounted for opacity in the $^{12}$CO line. Since we have a small sample size of 12, we place our estimates in context by comparing them to the catalogue presented in \cite{wu}, which shows 391 high-velocity molecular outflows from various sources in different evolutionary stages. The larger catalogue contains sources that are both low and high mass protostars.  We plot our estimated values along with the values calculated by \citep{wu} (Figure \ref{wu}). Specifically, we compare with the \cite{wu} results for (a) outflow mass vs energy and (b) IR luminosity of the central sources vs outflow energy.  Both plots show significant correlations, but this can be primarily attributed to all the axes scaling with $d^2$, where $d$ is the distance to the source.
%Even so, we still may expect that the most massive outflows are also the most energetic ones (panel a) and that outflow energy increases with the protostellar luminosity since both are proportional to accretion rate.  
In comparing the \cite{wu} data with our two sets of our results (i.e., estimates from all lines and those from $^{12}$CO alone), we see that both sets of estimates follow the general trends and scales from the population as a whole.  Furthermore, the margin between the $^{12}\mathrm{CO}$-only estimates and the multi-line estimates (Table \ref{12-13}) is small compared to the distribution of the broader population. 

The similarity of the distribution of both sets of estimated values to the larger population of outflows indicates our estimates and regression model are providing good estimates of outflow properties suitable for survey analysis. Overall, we estimate that the projected outflow properties have a 0.3~dex uncertainty and the unknown inclination of the angle suggests a further factor of 2 underestimate for the momentum and a factor of 2 underestimate for the energy assuming a uniform distribution of angles on the sky.

%------------------------------------------------------------------------------------------------------------
%%%%%%%%%%%%%%%%%%%%%%%%%%%%%%%%%%% CONCLUSIONS %%%%%%%%%%%%%%%%%%%%%%%%%%%%%%%%%%%%%%%%%%%%%%%%%%%%%%%%%%%%%%%%%%
\section{Conclusions}
In this paper, we have studied 13 molecular outflows in the Cygnus X region identified by \citepalias{gott12}, using JCMT observations of the $^{12}$CO(3-2), $^{13}$CO(3-2), and CO$^{18}$(3-2) spectral lines. We have calculated various properties of the outflows, identified associated infrared sources, and evaluated a new method to estimate gas column density from $^{12}$CO(3-2) line alone. 
\begin{enumerate}
    \item We present each of 13 molecular outflows in an atlas, displaying the extent of bipolarity, spatial and spectral extent of outflowing gas, along with the velocity distribution in PV-slices. All outflows except for G80.314+1.330 appear to be associated with clouds in the Cygnus X region.  The outflow G80.314+1.330 has a relatively larger negative LSR velocity and is likely associated with the Perseus Arm.
    
    \item Assuming LTE and uniform excitation temperature among the three CO lines we estimate mass, momentum, and energy of the remaining 12 outflows by following the method descried in \citepalias{deb18}. The results are summarized in Table \ref{prop_compare}. Our estimated values are comparable with those of a larger population study of outflows (\cite{wu}) and shown in Figure \ref{wu}. In particular, we find the mechanical luminosity of the outflows is $L_\mathrm{mech}\sim 10^{-3}L_{\mathrm{IR}}$.
    
    \item We also test a method of estimating of outflow properties from only $^{12}$CO(3-2) line data. We compare our $^{12}$CO(3-2)-only estimates with the three-line estimates. A relatively small but consistent underestimation (0.3 dex) is present in all three properties (mass, momentum, and energy; Figure \ref{corr}) and is likely due to projected linewidth of $^{13}$CO(3-2) being larger than the observed $^{13}$CO(3-2) line width so less emission is included in the outflow wings.  Since our sample is small, we compare the values with a compilation of properties from \citet{wu}. In this context, the outflow properties we measure are consistent with the general population and the uncertainties are within the scatter in the broader population (Figure \ref{wu}).
\end{enumerate}

After comparing the projected and estimated outflow properties we conclude that our $^{12}$CO-only optical depth model produces a fairly close correlation between estimated and projected values. Therefore we can utilize this model in our next work which will present a large survey of outflows in Cygnus X.  

%%%%%%%%%%%%%%%%%%%%%%%%%%%%%%%%%%% Acknowledgements %%%%%%%%%%%%%%%%%%%%%%%%%%%%%%%%%%%%%%%%%%%%%%%%%%%%%%%%%%%%%%%%%%
\section*{Acknowledgements}

The James Clerk Maxwell Telescope has historically been operated by the Joint Astronomy Centre on behalf of the Science and Technology Facilities Council of the United Kingdom, the National Research Council of Canada and the Netherlands Organisation for Scientific Research. The authors wish to recognize and acknowledge the significant cultural role and reverence that the summit of Maunakea has always had within the indigenous Hawaiian community.  We are most fortunate to have the opportunity to conduct observations from this mountain.  The authors acknowledge support from the Natural Sciences and Engineering Research Council of Canada, funding reference numbers RGPIN-2017-03987 and RGPIN 418517.

{\it Data Availability}: The data underlying this article are available in the Canadian Astronomy Data Centre, at https://dx.doi.org/10.11570/21.0001. \footnote{Data hosted during review at \url{https://www.canfar.net/storage/list/eros/OUTFLOWS_FITS}}

%%%%%%%%%%%%%%%%%%%%%%%%%%%%%%%%%%% REFERENCES  %%%%%%%%%%%%%%%%%%%%%%%%%%%%%%%%%%%%%%%%%%%%%%%%%%%%%%%%%%%%%%%%%%

%%%%%%%%%%%%%%%%% APPENDICES %%%%%%%%%%%%%%%%%%%%%

%%%%%%%%%%%%%%%%%%%%%%%%%%%%%%%%%%%%%%%%%%%%%%%%%%

\bsp	% typesetting comment
\label{lastpage}
\end{document}

\begin{center}
    {\bf Supplementary material for {\it Characterizing Outflows in the Cygnus X Region} by Deb et al.}
\end{center}

\section{Atlas of Molecular Outflows in Cygnus X}
\label{app:atlas}
\begin{figure*}[h!]
    \centering
    \includegraphics[width=0.95\textwidth]{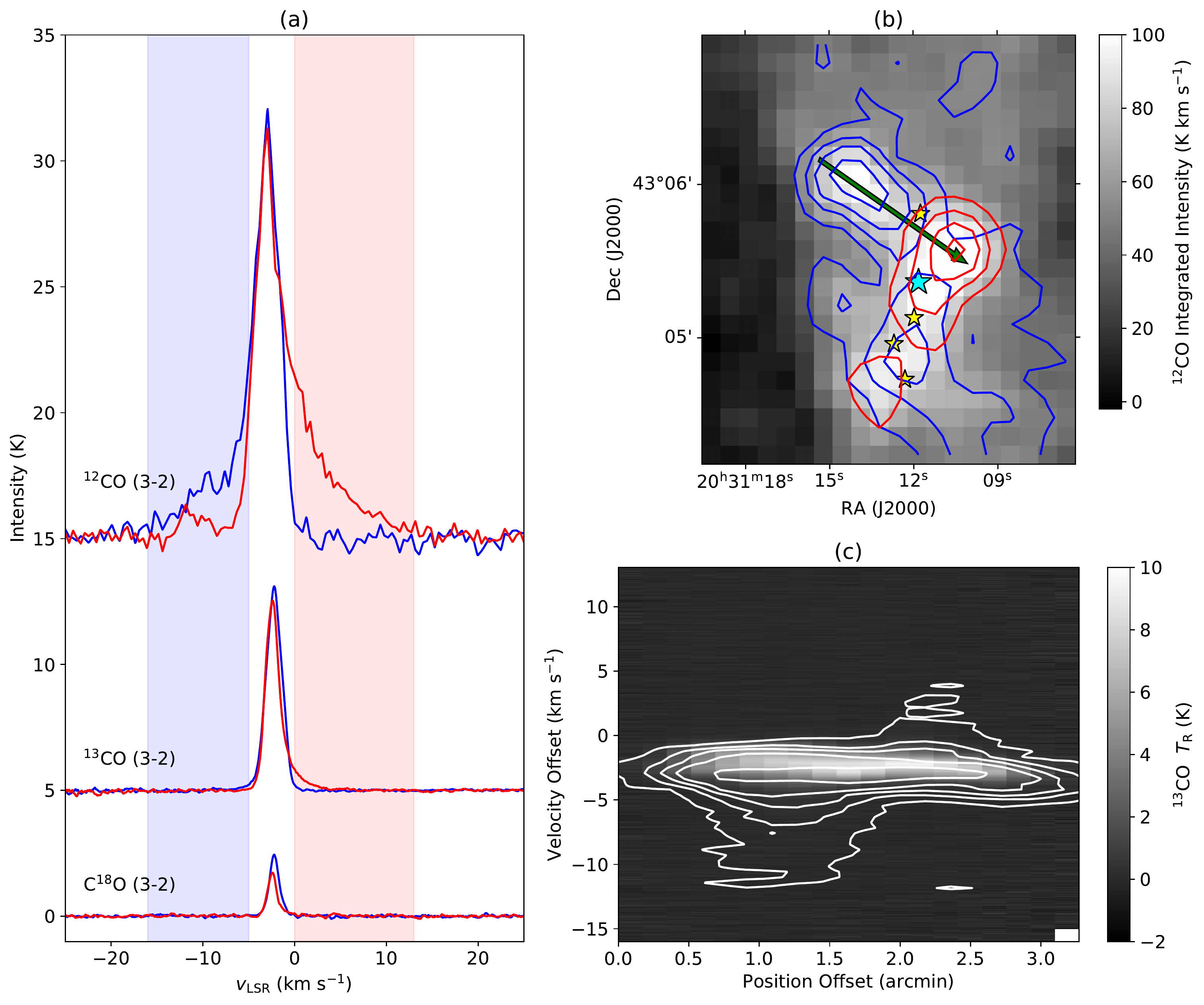}
    \caption{Outflow G81.435+2.147: (a) Blue- and redshifted outflow regions are shown in $^{12}$CO(3-2) (offset $+15$~K), $^{13}$CO(3-2) (offset $+5$~K), and C$^{18}$O(3-2) lines (b) Blue and red contour lines are obtained by integrating over velocity ranges from $v=-16$ to $-5\mathrm{~km~s}^{-1}$ and $v=0$ to $13\mathrm{~km~s}^{-1}$, and drawn at levels (7, 13, 20, 30, 40, 50) K~km~s$^{-1}$ and (10, 20, 30, 40, 50) K~km~s$^{-1}$ respectively. (c) Contours are drawn at levels (2, 5, 7.5, 10, 15, 20) K.}
    \label{of4u}
\end{figure*}

\begin{figure*}
    \centering
    \includegraphics[width=0.95\textwidth]{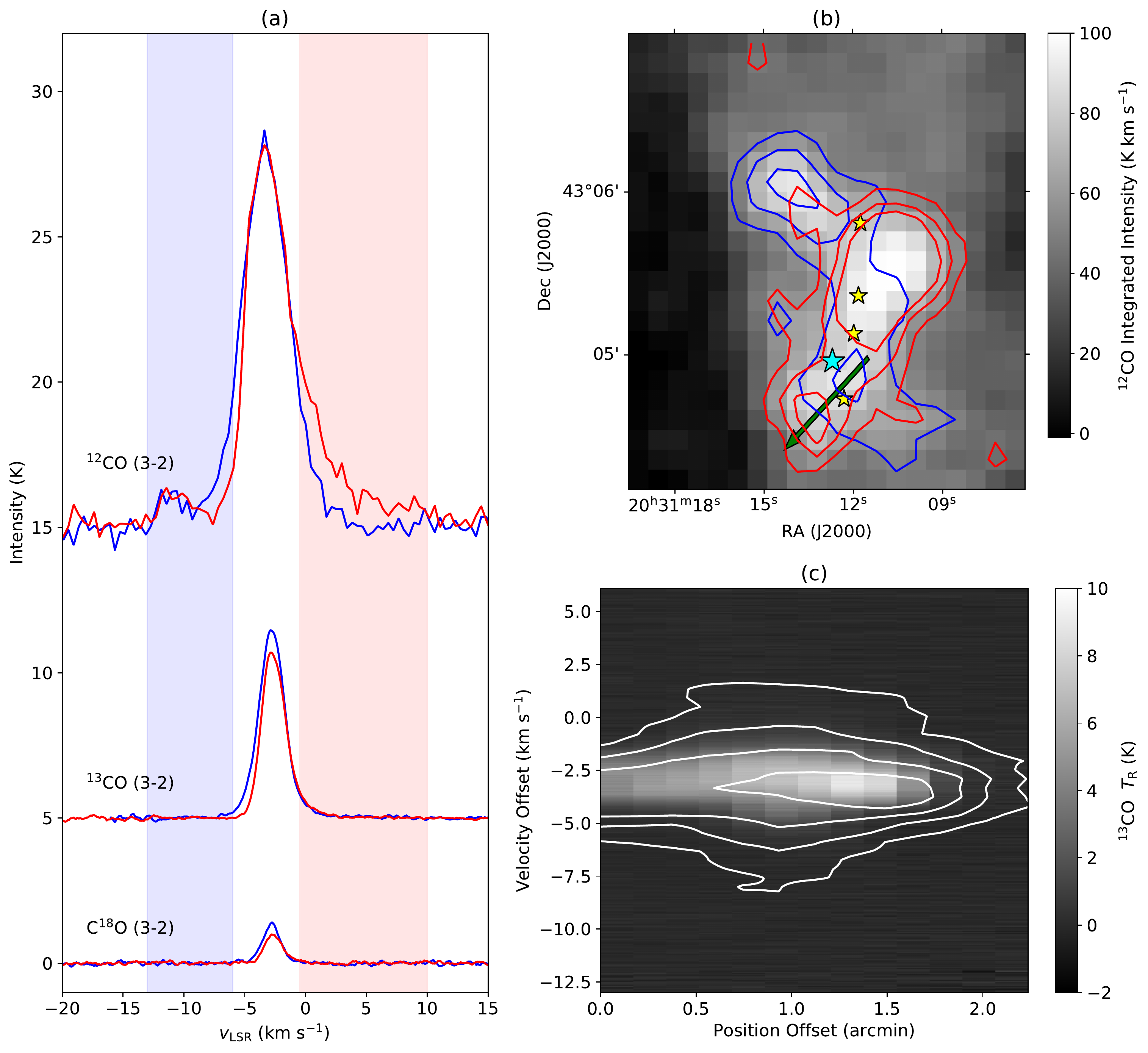}
    \caption{Outflow G81.424+2.140: (a) Blue- and redshifted outflow regions are shown in $^{12}$CO(3-2) (offset $+15$~K), $^{13}$CO(3-2) (offset $+5$~K), and C$^{18}$O(3-2) lines (b) Blue and red contour lines are obtained by integrating over velocity ranges from $v=-14$ to $-6\mathrm{~km~s}^{-1}$ and $v=-1.5$ to $6\mathrm{~km~s}^{-1}$, and drawn at levels (6, 13, 22) K~km~s$^{-1}$ and (4, 10, 24) K~km~s$^{-1}$ respectively. (c) Contours are drawn at levels (4, 10, 16) K.}
    \label{of4l}
\end{figure*}

\begin{figure*}
    \centering
    \includegraphics[width=0.95\textwidth]{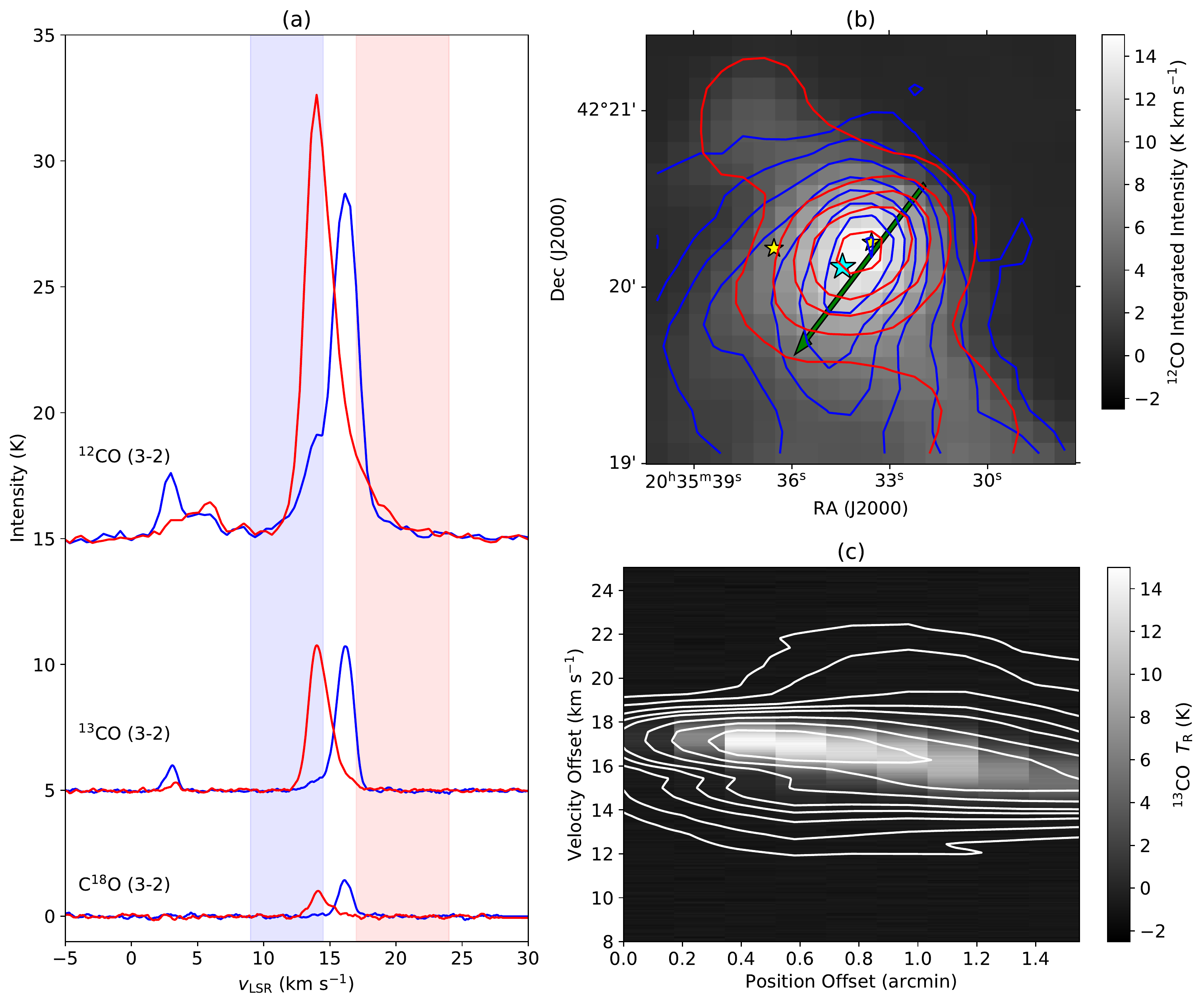}
    \caption{Outflow G81.302+1.055: (a) Blue- and redshifted outflow regions are shown in $^{12}$CO(3-2) (offset $+15$~K), $^{13}$CO(3-2) (offset $+5$~K), and C$^{18}$O(3-2) lines (b) Blue and red contour lines are obtained by integrating over velocity ranges from $v=9$ to $14.5\mathrm{~km~s}^{-1}$ and $v=17$ to $24\mathrm{~km~s}^{-1}$, and drawn at levels (0.45, 1, 2.5, 5, 7.5, 10, 12.5, 15, 20, 25, 28) K~km~s$^{-1}$ and (1, 3, 5, 7, 10, 15, 23) K~km~s$^{-1}$ respectively. (c) Contours are drawn at levels (1.2, 2.5, 5, 7.5, 10, 15, 20, 25) K.}
    \label{of5}
\end{figure*}

\begin{figure*}
    \centering
    \includegraphics[width=0.95\textwidth]{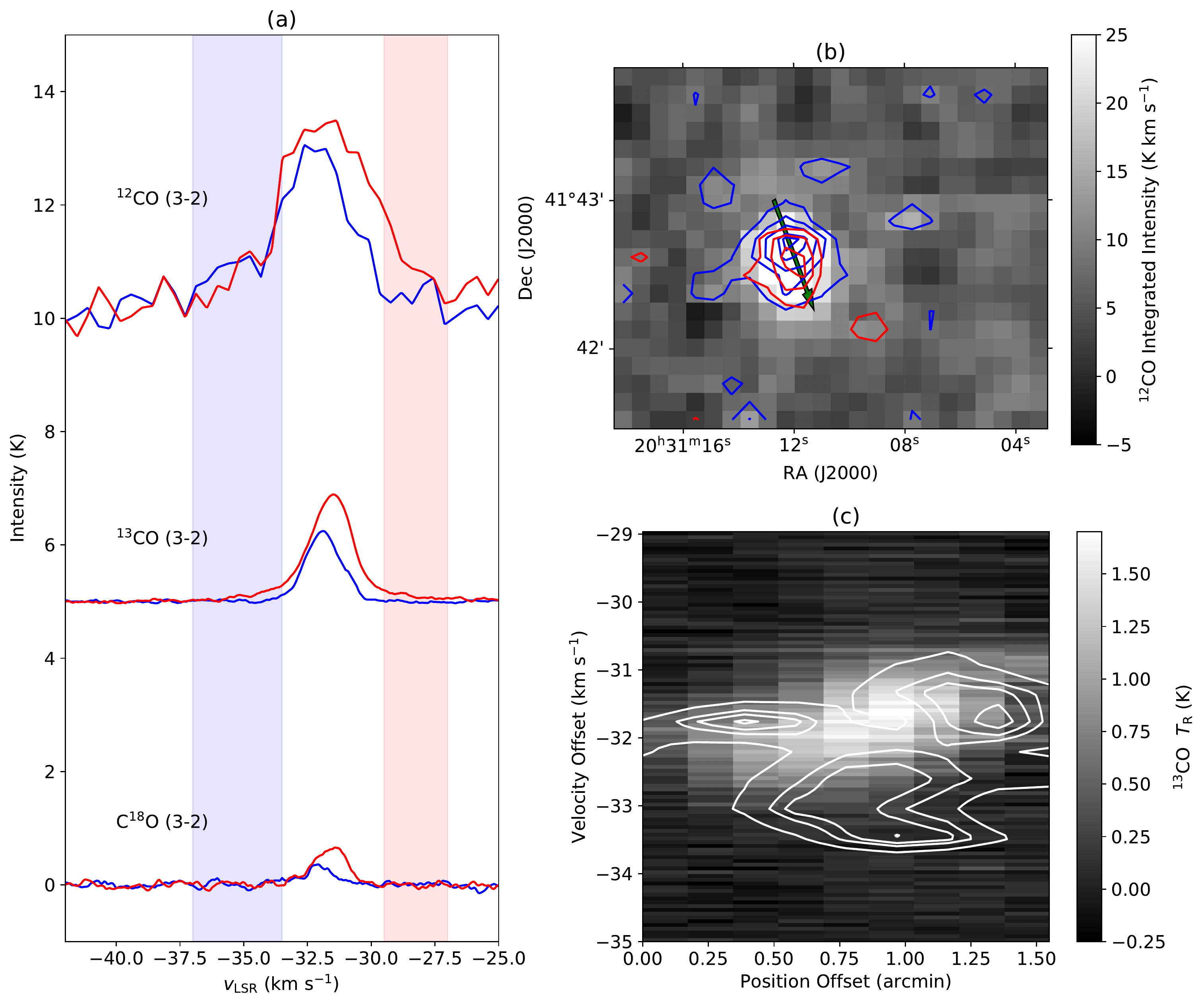}
    \caption{Outflow G81.424+2.140: (a) Blue- and redshifted outflow regions are shown in $^{12}$CO(3-2) (offset $+10$~K), $^{13}$CO(3-2) (offset $+5$~K), and C$^{18}$O(3-2) lines (b) Blue and red contour lines are obtained by integrating over velocity ranges from $v=-37$ to $-33.5\mathrm{~km~s}^{-1}$ and $v-29.5$ to $-27\mathrm{~km~s}^{-1}$, and drawn at levels (3, 7, 11, 16, 18) K~km~s$^{-1}$ and (3, 5, 7) K~km~s$^{-1}$ respectively. (c) Contours are drawn at levels (2, 2.6, 2.9, 3.3) K.}
    \label{of7}
\end{figure*}

\begin{figure*}
    \centering
    \includegraphics[width=0.95\textwidth]{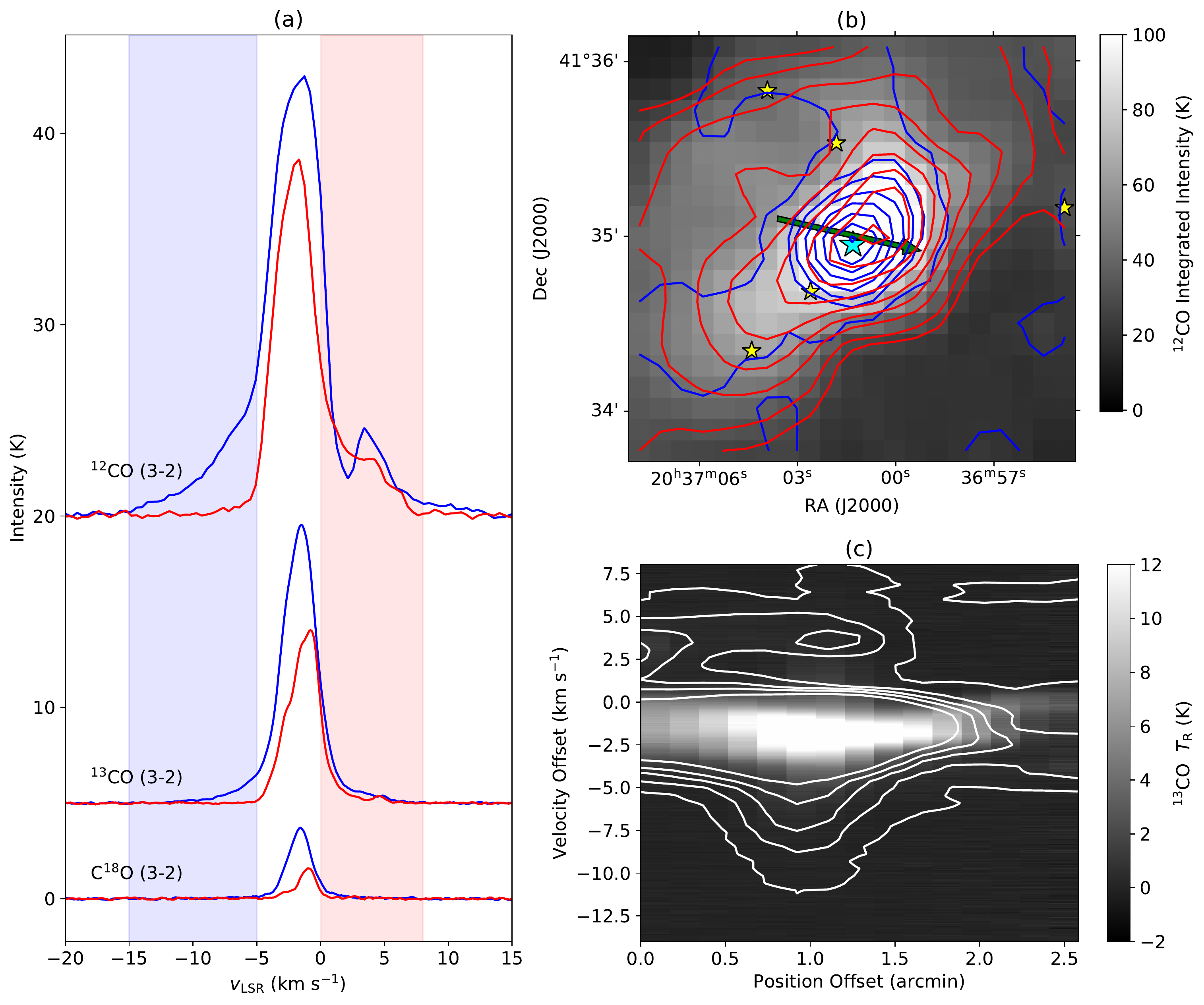}
    \caption{Outflow G80.862+0.385: (a) Blue- and redshifted outflow regions are shown in $^{12}$CO(3-2) (offset $+20$~K), $^{13}$CO(3-2) (offset $+5$~K), and C$^{18}$O(3-2) lines (b) Blue and red contour lines are obtained by integrating over velocity ranges from $v=-15$ to $-5\mathrm{~km~s}^{-1}$ and $v=0$ to $8\mathrm{~km~s}^{-1}$, and drawn at levels (7, 12, 16, 22, 30, 40, 50, 63) K~km~s$^{-1}$ and (15, 20, 30, 40, 50, 60, 70, 80, 90) K~km~s$^{-1}$ respectively. (c) Contours are drawn at levels (1.2, 3, 5, 7, 10) K.}
    \label{of10}
\end{figure*}

\begin{figure*}
    \centering
    \includegraphics[width=0.95\textwidth]{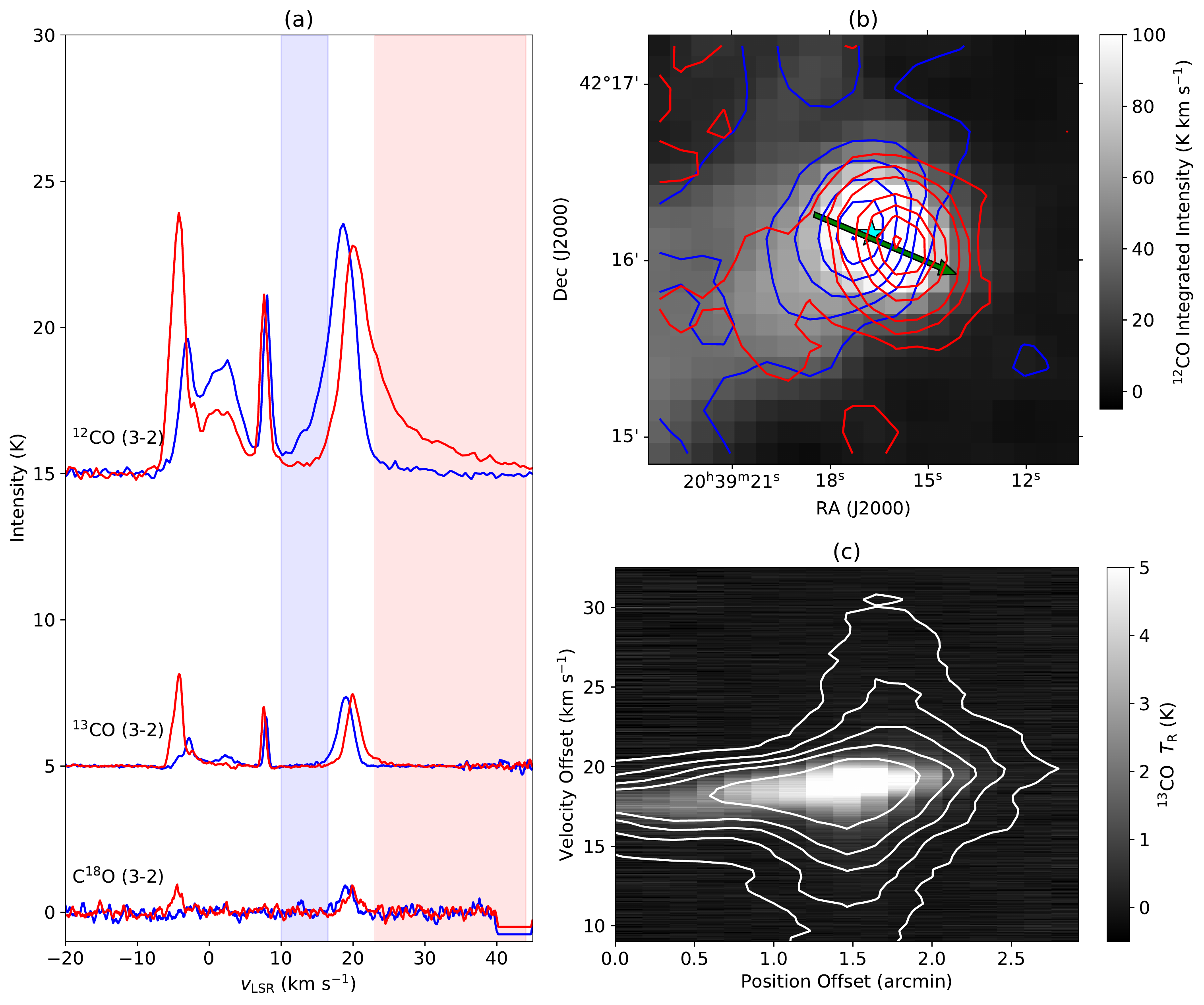}
    \caption{Outflow G81.663+0.468: (a) Blue- and redshifted outflow regions are shown in $^{12}$CO(3-2) (offset $+15$~K), $^{13}$CO(3-2) (offset $+5$~K), and C$^{18}$O(3-2) lines (b) Blue and red contour lines are obtained by integrating over velocity ranges from $v=10$ to $16.5\mathrm{~km~s}^{-1}$ and $v=23$ to $44\mathrm{~km~s}^{-1}$, and drawn at levels (3, 7, 12, 20, 30, 40) K~km~s$^{-1}$ and (1.5, 5, 10, 25, 40, 55, 75) K~km~s$^{-1}$ respectively. (c) Contours are drawn at levels (1.2, 3, 5, 7, 10) K.}
    \label{of14}
\end{figure*}

\begin{figure*}
    \centering
    \includegraphics[width=0.95\textwidth]{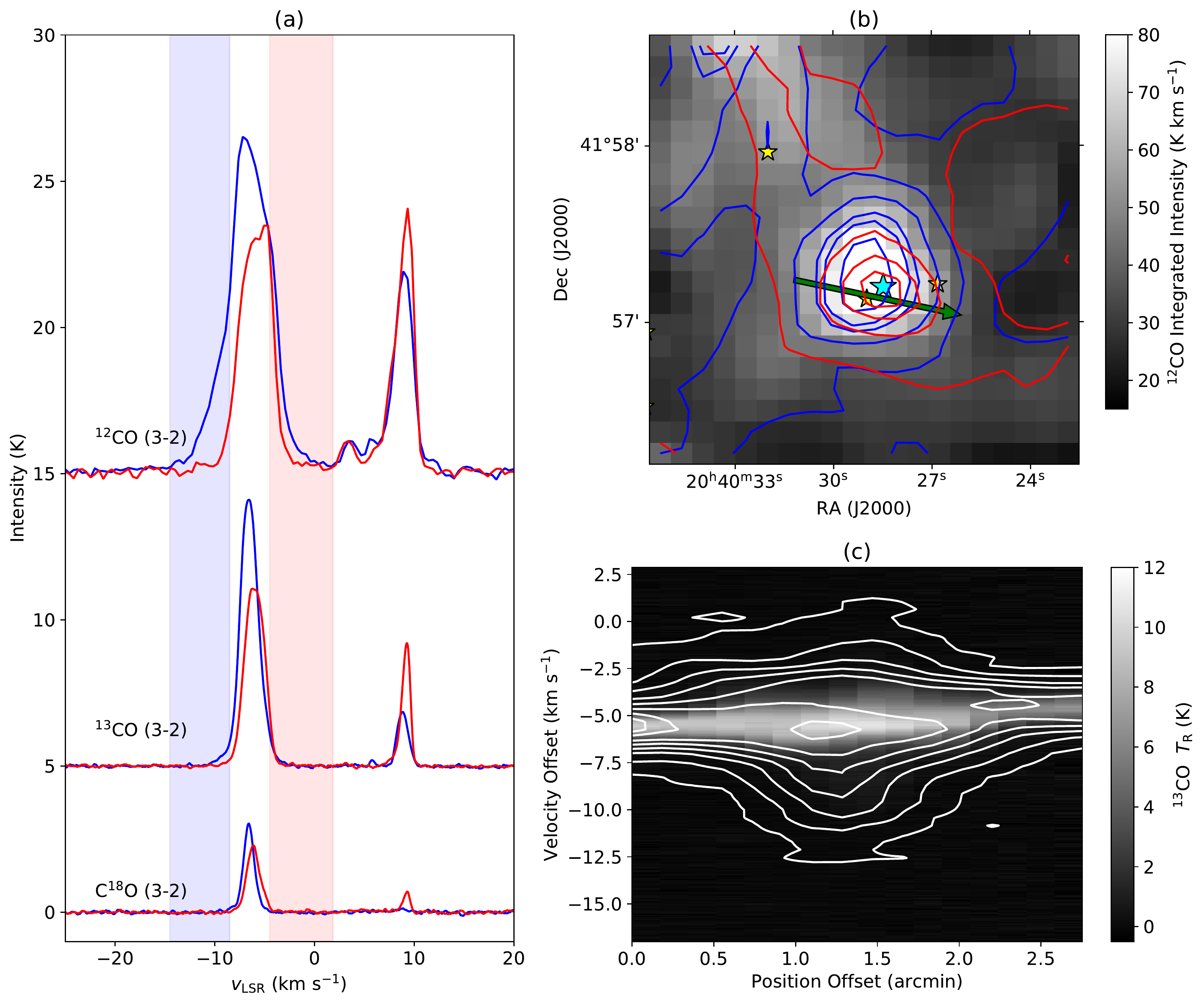}
    \caption{Outflow G81.551+0.098: (a) Blue- and redshifted outflow regions are shown in $^{12}$CO(3-2) (offset $+15$~K), $^{13}$CO(3-2) (offset $+5$~K), and C$^{18}$O(3-2) lines (b) Blue and red contour lines are obtained by integrating over velocity ranges from $v=-14.5$ to $-8.5\mathrm{~km~s}^{-1}$ and $v=-4.5$ to $1.8\mathrm{~km~s}^{-1}$, and drawn at levels (4, 12, 20, 28, 32, 40) K~km~s$^{-1}$ and (1, 8, 18, 25, 30, 35) K~km~s$^{-1}$ respectively. (c) Contours are drawn at levels (0.6, 2, 4, 6, 8, 10, 12, 13, 14) K.}
    \label{of16}
\end{figure*}

\begin{figure*}
    \centering
    \includegraphics[width=0.95\textwidth]{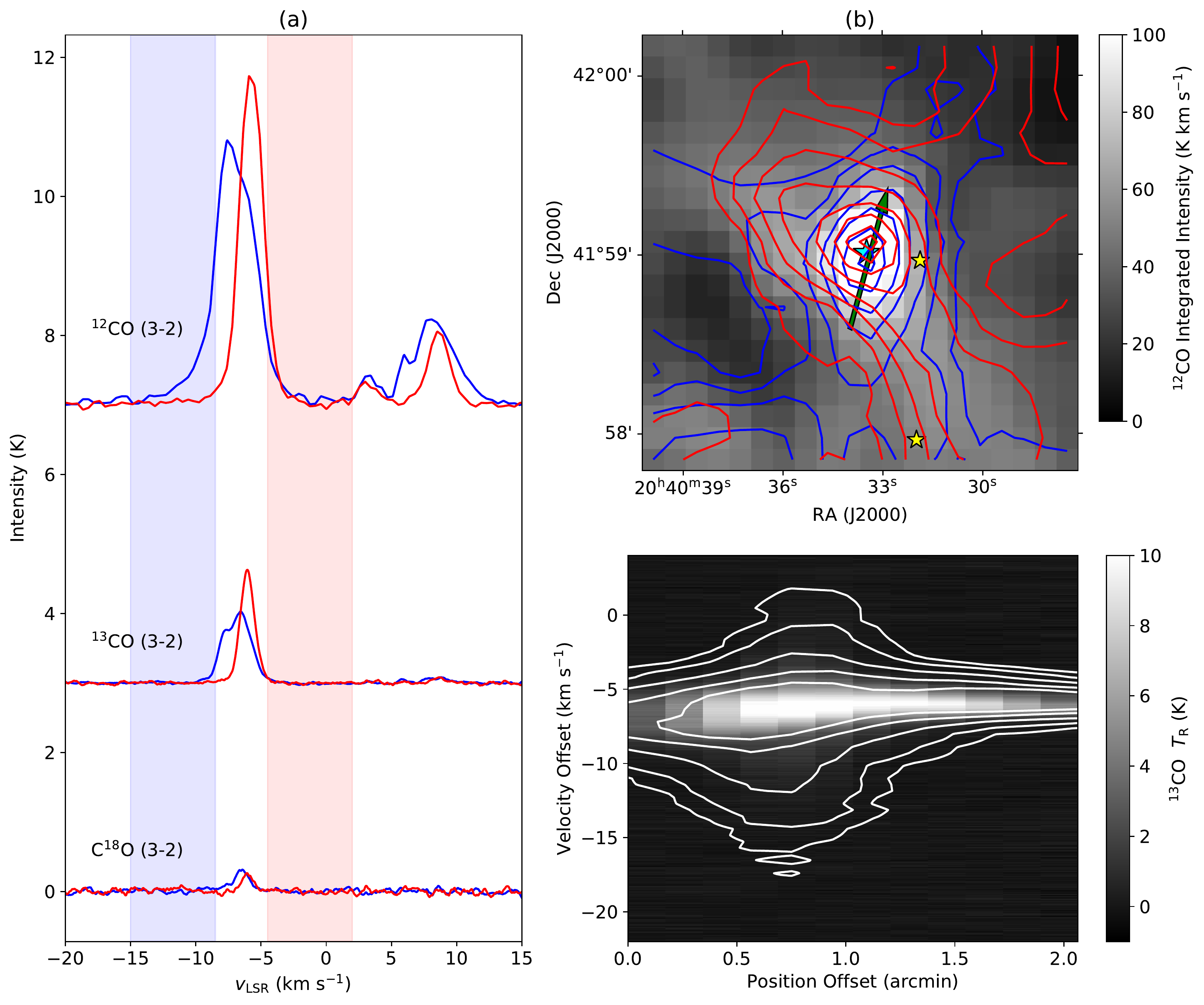}
    \caption{Outflow G81.582+0.104: (a) Blue- and redshifted outflow regions are shown in $^{12}$CO(3-2) (offset $+7$~K), $^{13}$CO(3-2) (offset $+3$~K), and C$^{18}$O(3-2) lines (b) Blue and red contour lines are obtained by integrating over velocity ranges from $v=-15$ to $-8.5\mathrm{~km~s}^{-1}$ and $v=-4.5$ to $2\mathrm{~km~s}^{-1}$, and drawn at levels (2.5, 4.5, 8, 16, 25, 38, 48, 53) K~km~s$^{-1}$ and (3, 5, 8, 12, 18, 25, 31, 35) K~km~s$^{-1}$ respectively. (c) Contours are drawn at levels (1.7, 3, 6, 10, 13, 17, 21) K.}
    \label{of17}
\end{figure*}

\begin{figure*}
    \centering
    \includegraphics[width=0.95\textwidth]{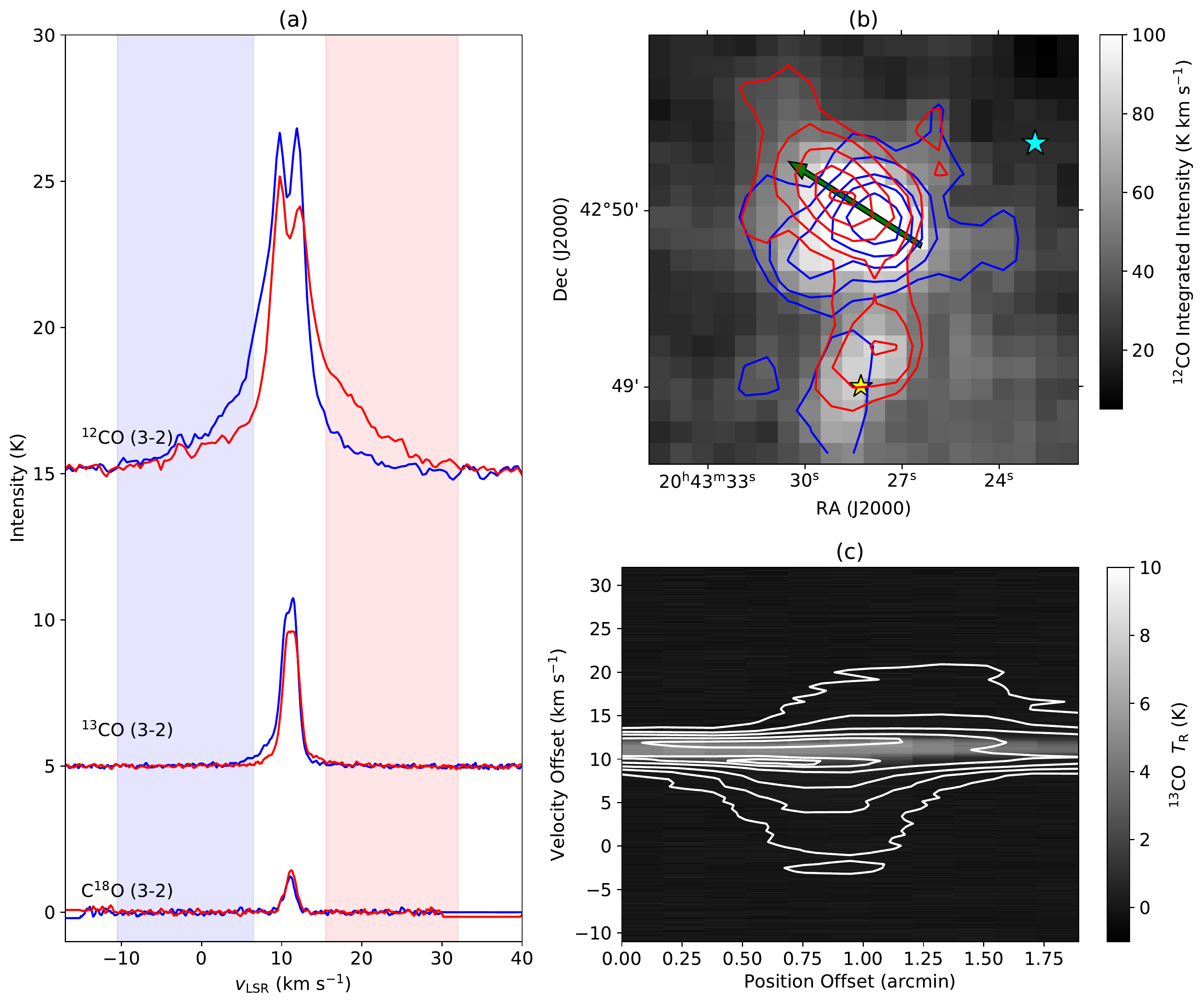}
    \caption{Outflow G82.581+0.203: (a) Blue- and redshifted outflow regions are shown in $^{12}$CO(3-2) (offset $+15$~K), $^{13}$CO(3-2) (offset $+5$~K), and C$^{18}$O(3-2) lines (b) Blue and red contour lines are obtained by integrating over velocity ranges from $v=-10.5$ to $6.5\mathrm{~km~s}^{-1}$ and $v=15.5$ to $32\mathrm{~km~s}^{-1}$, and drawn at levels (5, 10, 20, 30, 45, 68, 77) K~km~s$^{-1}$ and (5, 10, 20, 32, 42) K~km~s$^{-1}$ respectively. (c) Contours are drawn at levels (1.5, 3, 5, 8, 10, 11, 12) K.}
  \label{of19u}
\end{figure*}

\begin{figure*}
    \centering
    \includegraphics[width=0.95\textwidth]{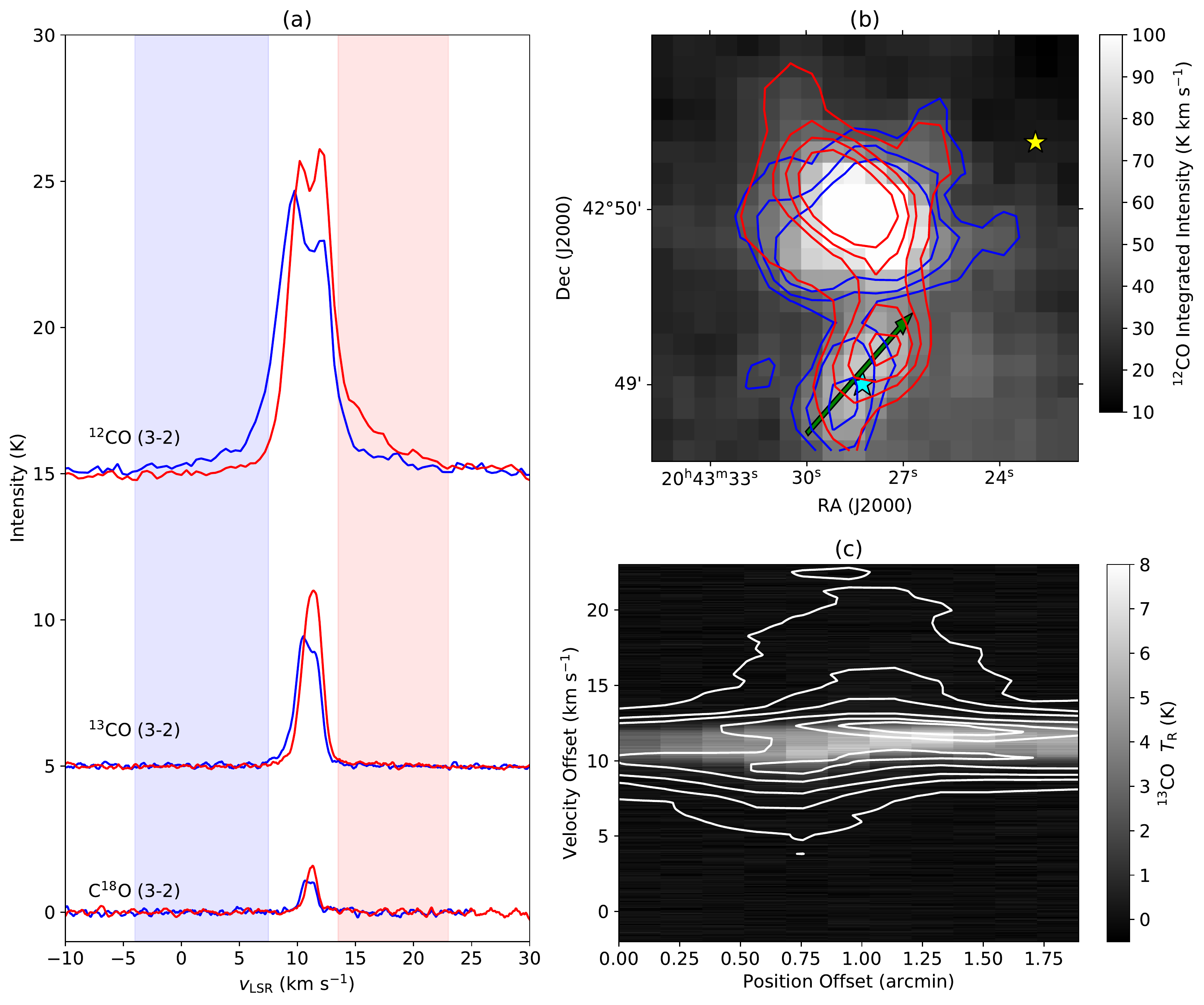}
    \caption{Outflow G82.571+0.194: (a) Blue- and redshifted outflow regions are shown in $^{12}$CO(3-2) (offset $+15$~K), $^{13}$CO(3-2) (offset $+5$~K), and C$^{18}$O(3-2) lines (b) Blue and red contour lines are obtained by integrating over velocity ranges from $v=-4$ to $7.5\mathrm{~km~s}^{-1}$ and $v=13.5$ to $23\mathrm{~km~s}^{-1}$, and drawn at levels (4, 7, 12) K~km~s$^{-1}$ and (5, 10, 15, 22) K~km~s$^{-1}$ respectively. (c) Contours are drawn at levels (1, 3, 5, 8, 10, 11, 12) K.}
    \label{of19l}
\end{figure*}

\begin{figure*}
    \centering
    \includegraphics[width=0.95\textwidth]{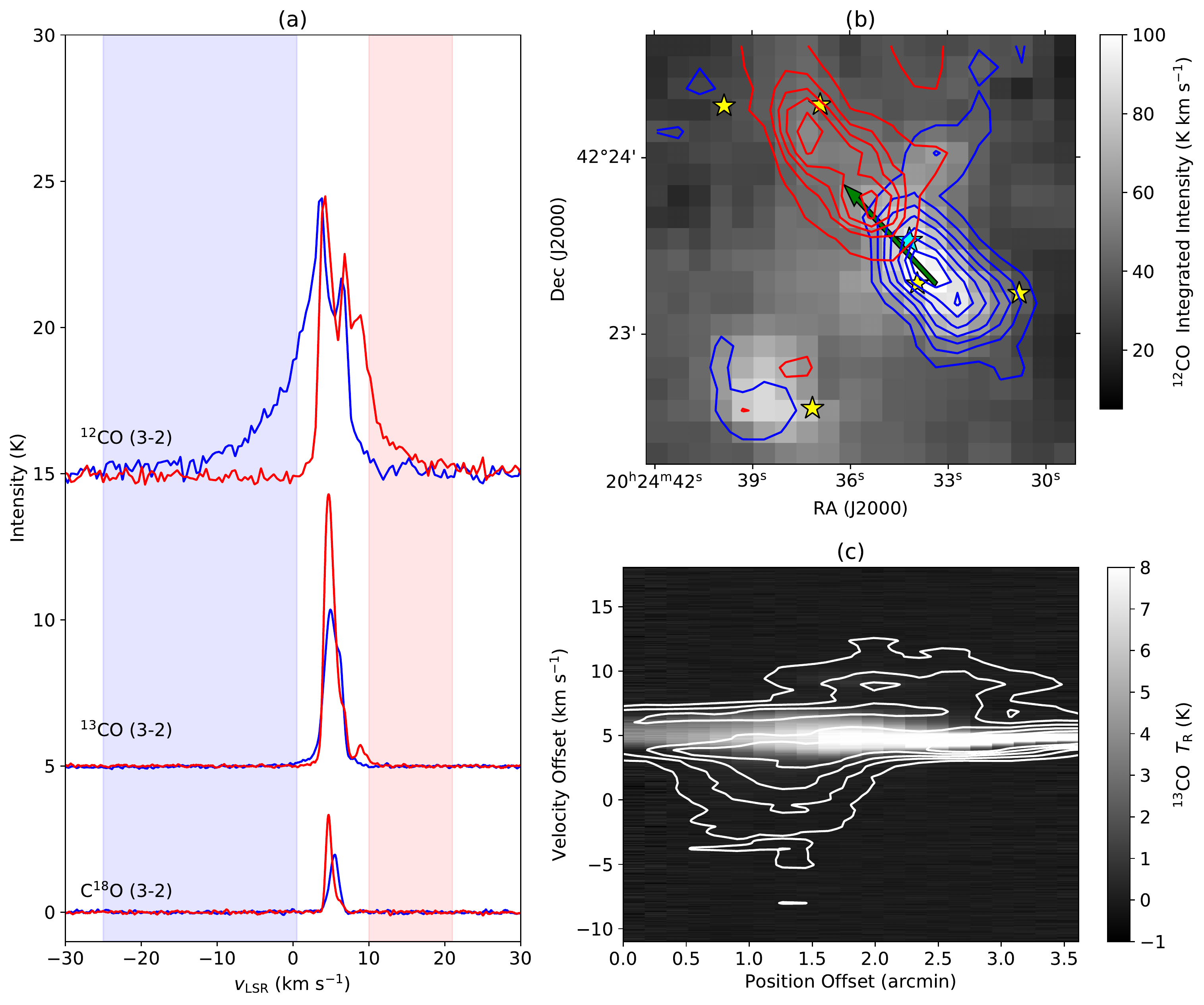}
    \caption{Outflow G80.158+2.727: (a) Blue- and redshifted outflow regions are shown in $^{12}$CO(3-2) (offset $+15$~K), $^{13}$CO(3-2) (offset $+5$~K), and C$^{18}$O(3-2) lines (b) Blue and red contour lines are obtained by integrating over velocity ranges from $v=-25$ to $0.5\mathrm{~km~s}^{-1}$ and $v=10$ to $21\mathrm{~km~s}^{-1}$, and drawn at levels (4, 9, 14, 19, 25, 30, 34) K~km~s$^{-1}$ and (5, 8, 10, 12, 15) K~km~s$^{-1}$ respectively. (c) Contours are drawn at levels (1.9, 4, 6, 8, 10) K.}
    \label{of23u}
\end{figure*}

\begin{figure*}
    \centering
    \includegraphics[width=0.95\textwidth]{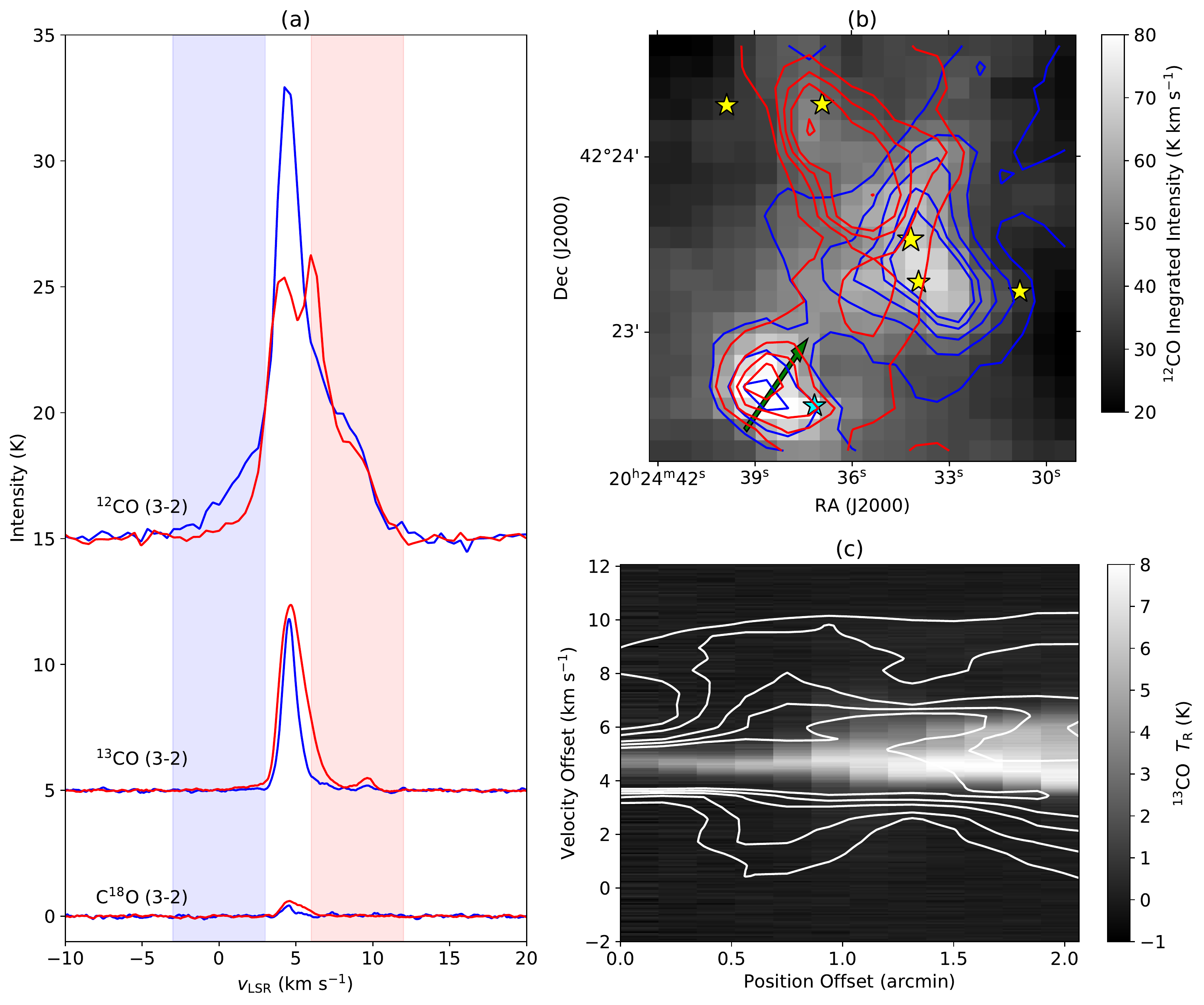}
    \caption{Outflow G80.149+2.710: (a) Blue- and redshifted outflow regions are shown in $^{12}$CO(3-2) (offset $+15$~K), $^{13}$CO(3-2) (offset $+5$~K), and C$^{18}$O(3-2) lines (b) Blue and red contour lines are obtained by integrating over velocity ranges from $v=-3$ to $3\mathrm{~km~s}^{-1}$ and $v=6$ to $12\mathrm{~km~s}^{-1}$, and drawn at levels (4, 9, 14, 19, 22) K~km~s$^{-1}$ and (8, 11, 13, 15, 25) K~km~s$^{-1}$ respectively. (c) Contours are drawn at levels (2, 3.5, 6, 8, 10) K.}
    \label{of23l}
\end{figure*}

Figures \ref{of4u} to \ref{of23l} show maps of the remaining 12 outflows analysed in the main text.

